\documentclass[slac_one,floatfix]{revtex4}
\usepackage{graphicx}
\usepackage{fancyhdr}
\newcommand{\mweak}{M_{\text{weak}}}
\newcommand{\msusy}{M_{\text{SUSY}}}
\newcommand{\mgut}{M_{\text{GUT}}}
\newcommand{\mplanck}{M_{\text{Pl}}}
\newcommand{\mstar}{M_{*}}
\newcommand{\Omegachi}{\Omega_{\chi}}
\newcommand{\OmegaM}{\Omega_{\text{M}}}
\newcommand{\OmegaLambda}{\Omega_{\Lambda}}
\newcommand{\OmegaDM}{\Omega_{\text{DM}}}
\newcommand{\OmegaB}{\Omega_B}
\newcommand{\Mpc}{\text{Mpc}}
\newcommand{\ifb}{\text{fb}^{-1}}

\newcommand{\gev}{\text{GeV}}
\newcommand{\tev}{\text{TeV}}

\newcommand{\s}{\text{s}}
\newcommand{\yr}{\text{yr}}

\newcommand{\eqref}[1]{Eq.~(\ref{#1})}

\newcommand{\secref}[1]{Sec.~\ref{sec:#1}}

\newcommand{\figref}[1]{Fig.~\ref{fig:#1}}
\newcommand{\figsref}[2]{Figs.~\ref{fig:#1} and \ref{fig:#2}}

\newcommand{\mchi}{m_{\chi}}

\newcommand{\gravitino}{\tilde{G}}

\newcommand{\mgaugino}{M_{1/2}}

\newcommand{\rem}[1]{{}}

\begin{document}

\title{
\begin{flushright}
\text{\small \rm UCI-TR-2005-34}
\end{flushright}
\vspace{12pt}
ILC Cosmology
\footnote{Plenary Colloquium presented at the 2005 International
Linear Collider Workshop, Stanford, California, USA, 18-22 March
2005.}  
} 

%

\author{Jonathan L.~Feng}
\affiliation{Department of Physics and Astronomy, University of
California, Irvine, CA 92697, USA}

\begin{abstract}
Recent breakthroughs in cosmology pose questions that require particle
physics answers.  I review the problems of dark matter, baryogenesis,
and dark energy and discuss how particle colliders, particularly the
International Linear Collider, may advance our understanding of the
contents and evolution of the Universe.
\end{abstract}

\maketitle

\pagestyle{plain}

\section{INTRODUCTION}

We are living through a period of scientific revolution: for the first
time in history, we have a compelling picture of the Universe on the
largest scales.  At the same time, we are preparing for a revolution
in particle physics as we explore the weak scale at the Tevatron, the
Large Hadron Collider (LHC), and the proposed International Linear
Collider (ILC).  Here I present one view of how these two revolutions
might be related.

\begin{figure*}[t]
\centering
\includegraphics[height=3.0in]{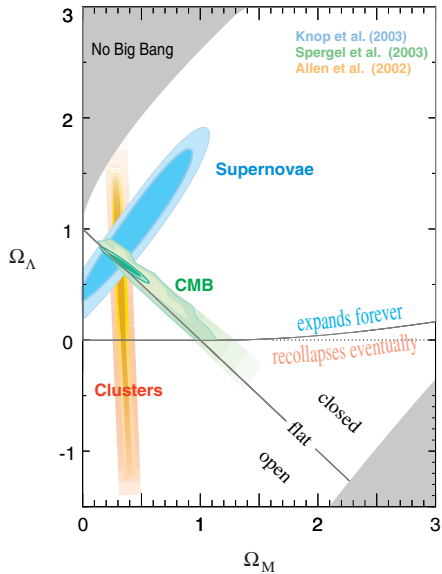}
\caption{Constraints on $\OmegaM$ and $\OmegaLambda$ from observations
  of supernovae, the CMB, and galaxy clusters~\cite{Knop:2003iy}.}
\label{fig:omegaplane}
\end{figure*}

In recent years, observations of supernovae, the cosmic microwave
background (CMB), and galaxy clusters have provided three stringent
constraints on $\OmegaM$ and $\OmegaLambda$, the energy densities of
matter and dark energy in units of the critical density.  These
results are consistent and favor $(\OmegaM, \OmegaLambda) \approx
(0.3, 0.7)$, as shown in \figref{omegaplane}.  The amount of matter in
the form of baryons is also constrained, both by the CMB and by the
observed abundances of light elements together with the theory of Big
Bang nucleosynthesis (BBN).  Although there are at present possibly
significant disagreements within the BBN data, the CMB and BBN data
taken as a whole are also impressively consistent, providing yet
another success for the current standard model of cosmology.

Through these and many other observations, the total energy densities
of non-baryonic dark matter, baryons, and dark energy are constrained
to be~\cite{Spergel:2003cb,Tegmark:2003ud}
\begin{eqnarray}
\OmegaDM &=& 23\% \pm 4\% \\
\OmegaB &=& 4\% \pm 0.4\% \\
\OmegaLambda &=& 73\% \pm 4\% \ .
\end{eqnarray}
These results are remarkable.  In particular, $\OmegaLambda$ is larger
than many expected, with profound implications for what the
cosmological constant problem is and how it might be solved. At least
two of the constraints of \figref{omegaplane} must be wrong to change
this conclusion.  These results are also remarkably precise --- the
fractional uncertainties on all three are ${\cal O}(10\%)$.  Given
that just a decade ago the range $0.2 \alt \OmegaDM \alt 0.6$ was
allowed and $\OmegaLambda =0$ was often assumed, this represents
spectacular progress.  Although much of cosmology remains imprecise,
as we will see, the quantum leap in precision in these three
quantities already has dramatic implications for particle physics.

At the same time, recent progress in cosmology is probably best viewed
as the first steps on the road to understanding the Universe.
Consider an historical precedent: in 200 B.C., Eratosthenes determined
the size of the Earth.  On a day when the Sun was directly overhead in
Syene, Eratosthenes sent a graduate student to measure the lengths of
shadows in Alexandria.  He was then able to extrapolate from the known
distance between these two cities to determine the circumference of
the Earth.  His answer was
\begin{equation}
2 \pi R_{\oplus} = 250,000~\text{stadia} \ .
\end{equation}
This result is remarkable.  At the time of publication, it was bigger
than many expected, leading many to be skeptical and helping to earn
Eratosthenes the nickname ``Beta''~\cite{Heath}.  His result was also
remarkably precise. We now know that it was good to less than
10\%~\cite{Goldstein,Rawlins1,Rawlins2}, where the leading source of
uncertainty is systematic error from the exact definition of the unit
``stadion''~\cite{Gulbekian}.  At the same time, the achievement of
Eratosthenes, though important, could hardly be characterized as a
complete understanding of the Earth.  Rather, it was just the
beginning of centuries of exploration, which eventually led to the
mapping of continents and oceans, giving us the picture of the Earth
we have today.

\begin{figure*}[tb]
\centering
\includegraphics[width=0.7\textwidth]{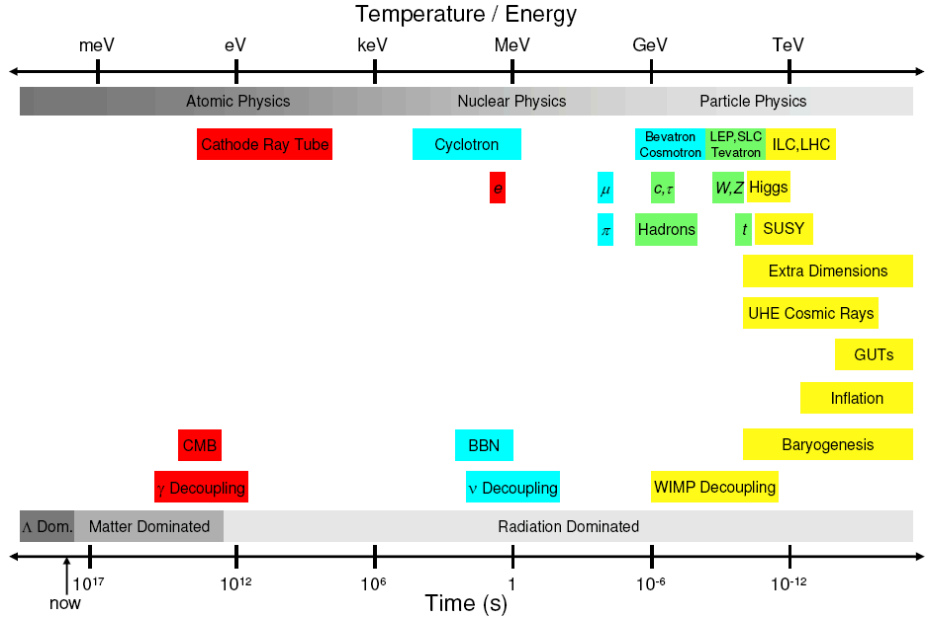}
\caption{Milestones in particle physics and cosmology along the
cosmological timeline.}
\label{fig:timeline}
\end{figure*}

In a similar vein, recent breakthroughs in cosmology answer many
questions, but highlight even more.  These include
\begin{itemize}
\item Dark Matter: What is it?  How is it distributed?  How does it
  impact structure formation?
\item Baryons: Why is there an asymmetry between matter and
  anti-matter?  Why does $\OmegaB$ have the value it has?
\item Dark Energy: What is it?  Why isn't $\OmegaLambda \sim 10^{120}$?
  Why isn't $\OmegaLambda$ zero?  How does it evolve?
\end{itemize}
Although these questions will continue to be sharpened by
astrophysical observations at large length scales, it is clear that
satisfying answers will require fundamental progress in our
understanding of microphysics.  This is nothing new --- the history of
advances in cosmology is to a large extent the story of successful
synergy between studies of the Universe on the smallest and largest
length scales.  This interplay is shown in \figref{timeline}, where
several milestones in particle physics and cosmology are placed along
the cosmological timeline.  As particle experiments reach smaller
length scales and higher energies, they probe times closer to the Big
Bang.  Just as atomic physics is required to interpret the CMB signal
from $t \sim 10^{13}~\s$ after the Big Bang and nuclear physics is
required to extrapolate back to BBN at $t \sim 1~\s$, particle
physics, and particularly the physics of the electroweak scale, is
required to understand the era before $t \sim 10^{-8}~\s$, the era
that contains the answers to many of our most basic questions.

In the next few years, the Tevatron and LHC will play the crucial role
of opening the door to the weak scale, to be followed, we hope, by
detailed studies of new physics at these colliders and the proposed
ILC.  Here I will give an overview of the potential role of particle
colliders, and especially the ILC, in answering cosmological
questions.  The discussion begins with dark matter, a subject in which
there are many concrete and compelling connections between the ILC and
cosmology. We will then consider baryogenesis and dark energy.  Much
of the work described here is an outgrowth of the activities of the
Cosmology Subgroup of the American Linear Collider Physics Group.
Additional background and results may be found in the Cosmology
Subgroup's report~\cite{whitepaper}, as well as in other contributions
to the proceedings of this workshop~\cite{Gray:2005ci,Steffen:2005cn,%
Mambrini:2005uw,Birkedal:2005jq,Kanemura:2005cj,deBoer:2005nf,%
Battaglia:2005ie}.

\section{DARK MATTER}

\subsection{Dark Matter and the Weak Scale}

The particle or particles that make up most of dark matter must be
stable, at least on cosmological time scales, and non-baryonic, so
that they do not disrupt the successes of BBN.  They must also be cold
or warm to properly seed structure formation, and their interactions
with normal matter must be weak enough to avoid violating current
bounds from dark matter searches.  The stringency of these criteria
pale in comparison with the unbridled enthusiasm of theorists, who
have proposed scores of viable candidates with masses and interaction
cross sections varying over tens of orders of magnitude.

Candidates with weak scale masses have received much of the attention,
however. There are at least four good reasons for this.  First, these
proposals are testable.  Second, new particles at the weak scale are
independently motivated by attempts to understand electroweak symmetry
breaking.  Third, these new particles often ``automatically'' have all
the right properties to be dark matter.  For example, their stability
often follows as a result of discrete symmetries that are necessary to
make electroweak theories viable, independent of cosmology.  And
fourth, these new particles are naturally produced with the
cosmological densities required of dark matter.

\begin{figure*}[t]
\centering
\includegraphics[height=2.4in]{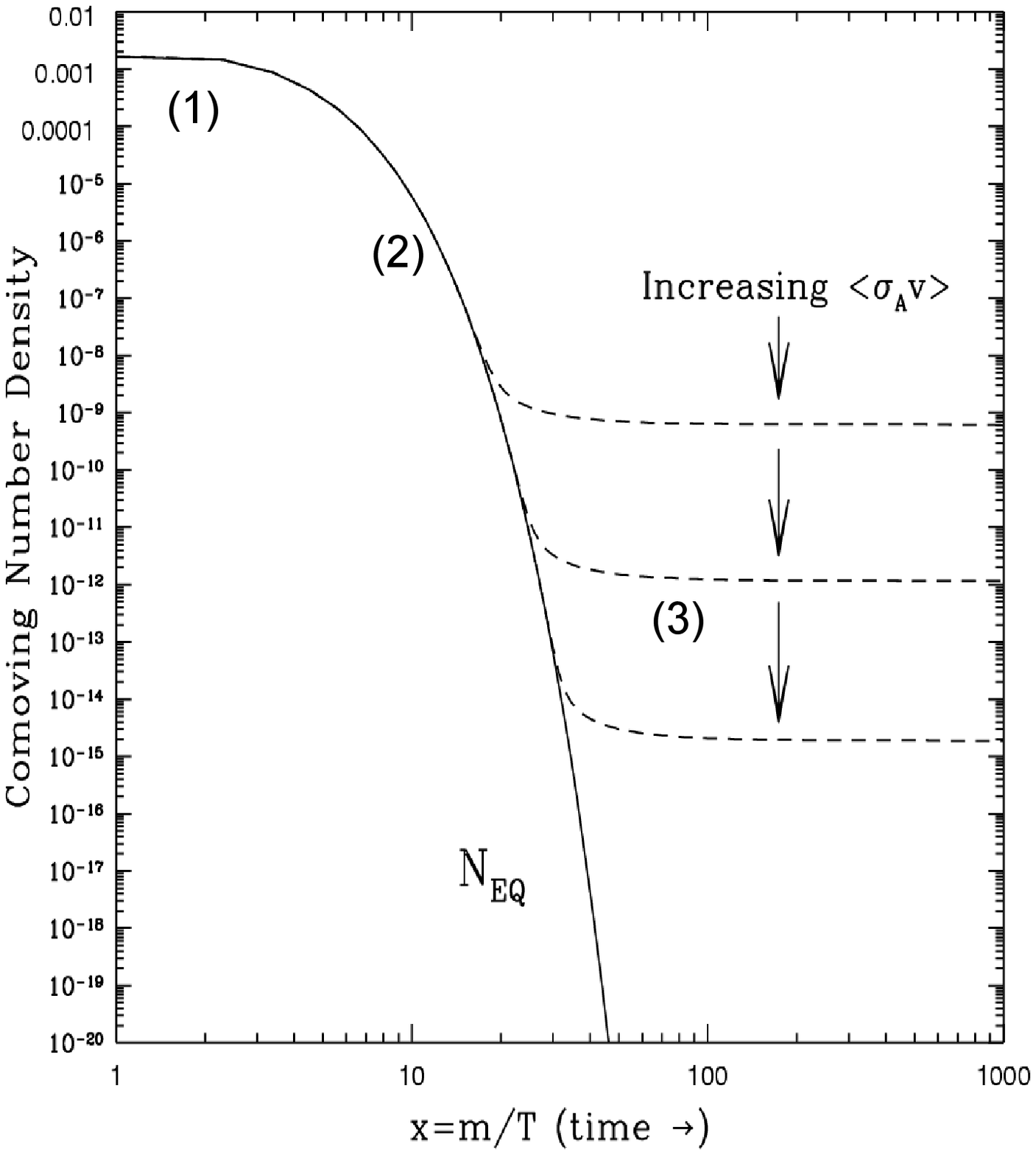}
\hspace*{0.5in}
\includegraphics[height=2.4in]{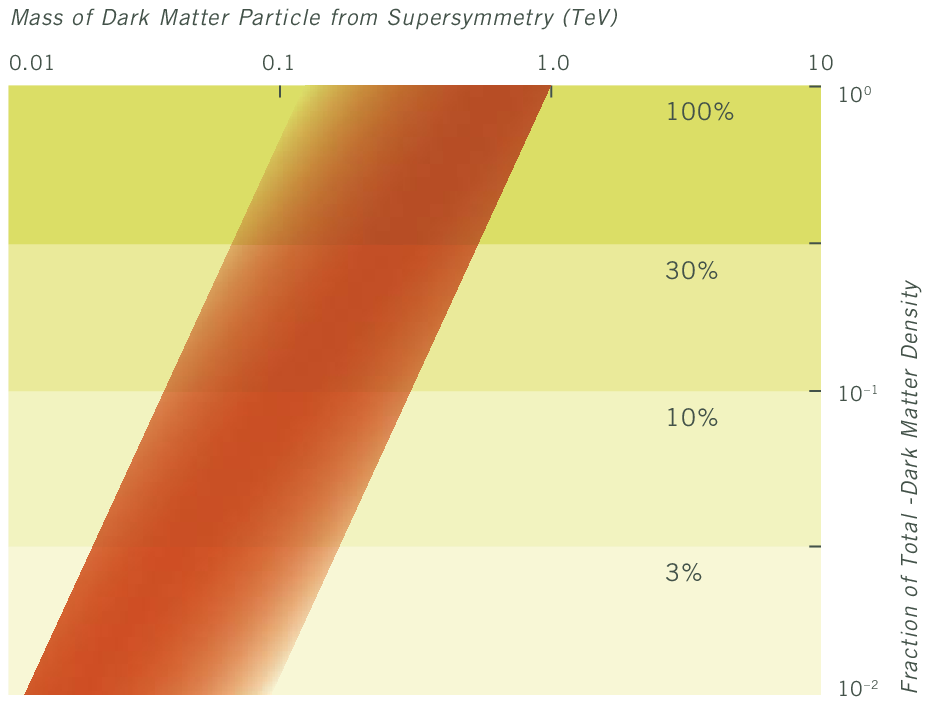}
\caption{Left: The cosmological evolution of a thermal relic's
  comoving number density~\cite{Jungman:1995df}. Right: A band of
  natural values in the $(\mchi, \Omegachi)$ plane for a thermal
  relic~\protect\cite{discovering}.}
\label{fig:freezeout}
\end{figure*}

The last motivation is particularly tantalizing.  Dark matter may be
produced in a simple and predictive manner as a thermal relic of the
Big Bang. The evolution of a thermal relic's number density is shown
in \figref{freezeout}. In stage (1), the early Universe is dense and
hot, and all particles are in thermal (chemical) equilibrium.  In
stage (2), the Universe cools to temperatures $T$ below the dark
matter particle's mass $\mchi$, and the number of dark matter
particles becomes Boltzmann suppressed, dropping exponentially as
$e^{-\mchi/T}$.  In stage (3), the Universe becomes so cool and dilute
that the dark matter annihilation rate is too low to maintain
equilibrium.  The dark matter particles then ``freeze out,'' with
their number asymptotically approaching a constant, their thermal
relic density.

More detailed analysis shows that the thermal relic density is rather
insensitive to $\mchi$ and inversely proportional to the annihilation
cross section: $\OmegaDM \sim \langle \sigma_A v \rangle^{-1}$.  The
constant of proportionality depends on the details of the
microphysics, but we may give a rough estimate.  On dimensional
grounds, the cross section can be written
\begin{equation}
\sigma_A v = k \frac{4 \pi \alpha_1^2}{m_{\chi}^2} \ 
(\text{1 or $v^2$})\ ,
\end{equation}
where $v$ is the relative velocity of the annihilating particles, the
factor $v^2$ is absent or present for $S$- or $P$-wave
annihilation, respectively, and terms higher-order in $v$ have been
neglected.  The constant $\alpha_1$ is the hypercharge fine structure
constant, and $k$ parameterizes deviations from this estimate.

With this parametrization, given a choice of $k$, the relic density is
determined as a function of $\mchi$.  The results are shown in
\figref{freezeout}.  The width of the band comes from considering both
$S$- and $P$-wave annihilation, and from letting $k$ vary from
$\frac{1}{2}$ to 2. We see that a particle that makes up all of dark
matter is predicted to have mass in the range $\mchi \sim 100~\gev -
1~\tev$; a particle that makes up 10\% of dark matter, still
significant with respect to its impact on structure formation, for
example, has mass $\mchi \sim 30~\gev - 300~\gev$.  There are models
in which the effective $k$ is outside our illustrative range.  In
fact, values of $k$ smaller than we have assumed, predicting smaller
$\mchi$, are not uncommon, as the masses of virtual particles in
annihilation diagrams can be significantly higher than $\mchi$.
However, the general conclusion remains: particles with mass at the
weak scale naturally have significant thermal relic densities.  For
this reason, even null results from LHC and ILC searches for dark
matter are important, and a thorough exploration of the weak scale
will play a crucial role in attempts to identify the particle or
particles that make up dark matter.

Given these results, many theories for new physics at the weak scale
contain promising dark matter candidates.  The candidates that exploit
the tantalizing numerical ``coincidence'' shown in \figref{freezeout}
may be grouped into two classes: WIMPs and
superWIMPs~\cite{Feng:2003zu}. In the following subsections, we
consider what insights colliders may provide in each of these two
cases.

\subsection{WIMPs}

Weakly-interacting massive particles (WIMPs) have weak-scale masses
and weak-scale interactions.  They include neutralinos in
supersymmetry~\cite{Goldberg:1983nd}, Kaluza-Klein particles in
theories with universal extra
dimensions~\cite{Servant:2002aq,Cheng:2002ej}, branons in theories
with large extra dimensions~\cite{Cembranos:2003mr}, and the lightest
$T$-odd particle in some little Higgs theories~\cite{Cheng:2003ju}.

The study of WIMP dark matter at colliders may be divided into three
(overlapping) stages:
\begin{enumerate}
\item WIMP Candidate Identification.  Is there evidence for WIMPs from
  colliders from, for example, events with missing energy and
  momentum?  What are the candidates' masses, spins, and other quantum
  numbers?
\item WIMP Relic Density Determination. What are the dark matter
  candidates' predicted thermal relic densities?  Can they be
  significant components or all of dark matter?
\item Mapping the WIMP Universe. Combined with results from direct and
  indirect dark matter searches, what can collider studies tell us
  about astrophysical questions, such as the distribution of dark
  matter in the Universe?
\end{enumerate}
Stage 1 is discussed in Ref.~\cite{whitepaper}.  In the following
subsections, we will explore how well the LHC and ILC may advance
Stages 2 and 3.

\begin{figure*}[t]
\centering
\includegraphics[height=2.5in]{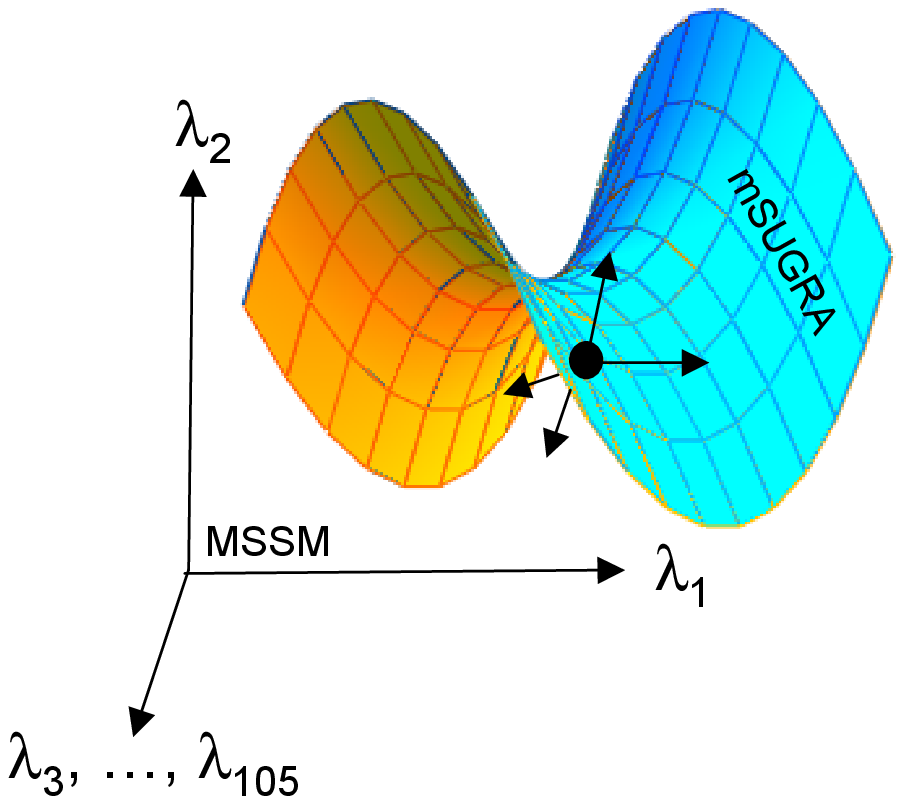}
\hspace*{0.5in}
\includegraphics[height=2.5in]{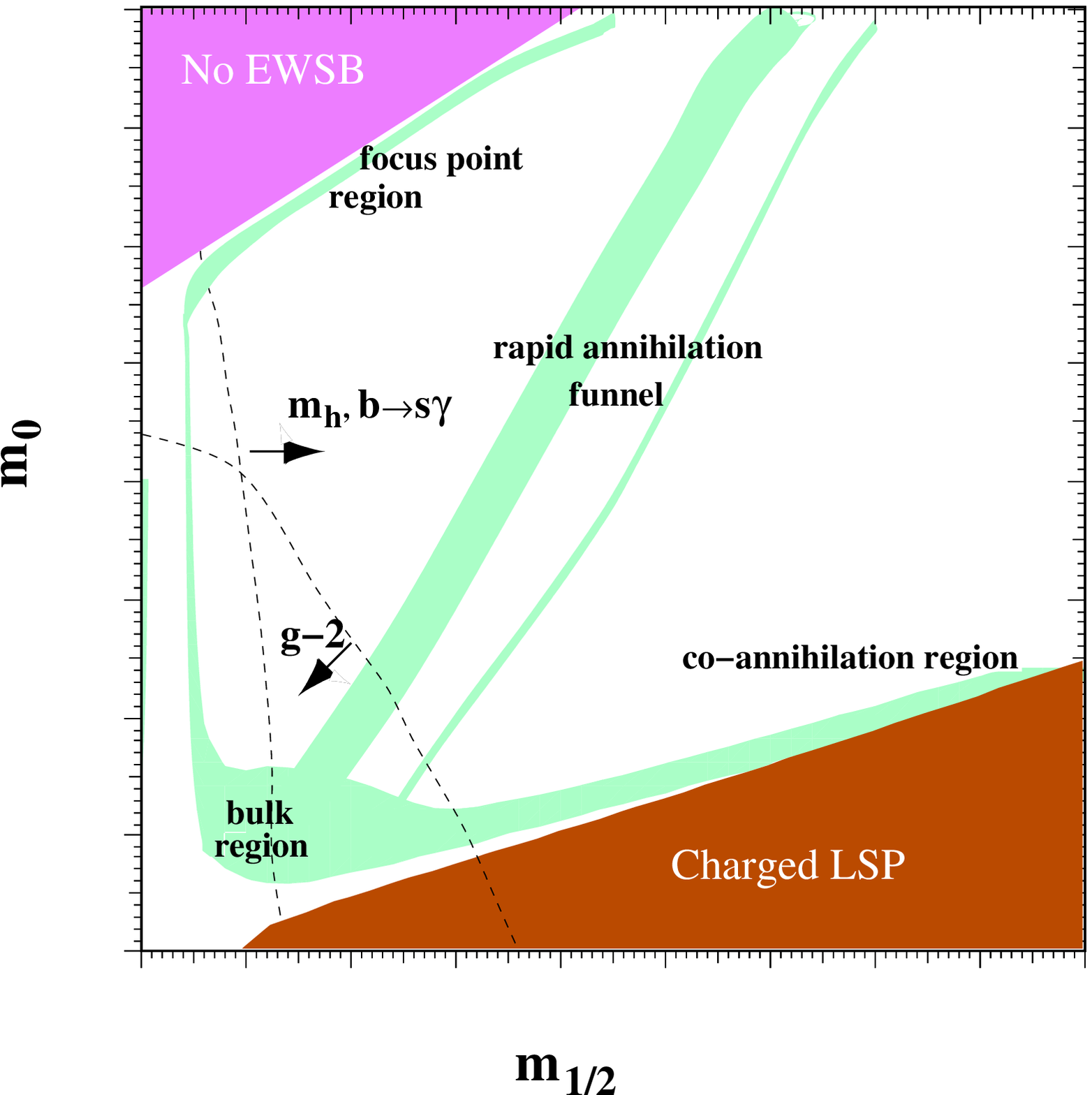}
\caption{Left: Minimal supergravity (mSUGRA) defines a hypersurface in
the 105-dimensional parameter space of the MSSM.  In the studies
described here, the underlying parameters are assumed to lie on the
mSUGRA hypersurface, but deviations in all directions of the MSSM
parameter space are allowed when evaluating the potential of colliders
to constrain parameters.  Right: Schematic diagram of regions with the
right amount of dark matter (shaded) in mSUGRA. This diagram is
qualitative.  The locations of the shaded regions depend on suppressed
parameters, and axis labels are purposely omitted~\cite{whitepaper}. }
\label{fig:parameters}
\end{figure*}

To address these issues concretely, it is necessary to focus on one
representative example, typically neutralinos. Even with this
restriction, there are many qualitatively different scenarios. A
common choice is to study minimal supergravity (mSUGRA), a simple model
framework that encompasses many different possibilities.  In this
case, one assumes that the underlying supersymmetry parameters
realized in Nature are those of a point in mSUGRA parameter space.  In
determining the capabilities of colliders, however, it is best to
relax all mSUGRA assumptions and ask how well the 105 parameters of
the general Minimal Supersymmetric Standard Model (MSSM) may be
determined.  This approach is illustrated in \figref{parameters}.

In much of mSUGRA parameter space the neutralino relic density lies
above the narrow allowed window, and so these possibilities are
cosmologically excluded.  The regions in which the neutralino relic
density is not too large, but is still sufficient to be all of dark
matter, are cosmologically preferred.  They have been given names and
include the bulk, focus point, co-annihilation, and rapid annihilation
funnel regions shown in \figref{parameters}.  Results from
representative models in each of the first two regions are summarized
below.  For results for the other two regions and related studies, see
Refs.~\cite{Gray:2005ci,Birkedal:2005jq,Battaglia:2005ie,%
Moroi:2005nc,Allanach:2004xn}.  For each model, the superpartner
spectrum is determined by ISAJET~\cite{Paige:2003mg}, and cosmological
observables, such as the thermal relic density, are determined by
DARKSUSY~\cite{Gondolo:2004sc} and micrOMEGAs~\cite{Belanger:2004yn}.

\subsubsection{WIMP Relic Density Determination \label{sec:relic}}

\paragraph{Bulk Region}

In the bulk region, a much studied model is specified by the mSUGRA
parameters
\begin{equation}
\text{LCC1:} \ (m_0, M_{1/2}, A_0, \tan\beta) = 
(100~\gev, 250~\gev, -100~\gev, 10) \quad
  [\mu > 0, m_{3/2} > m_{\text{LSP}}, m_t = 178~\gev] \ . 
\end{equation}
The neutralino thermal relic density at this point is $\Omegachi h^2 =
0.19$ ($h \simeq 0.71$), significantly higher than allowed by the
latest cosmological constraints.  Nevertheless, the choice of LCC1 is
convenient, since it has been studied in great detail in other
studies, where it is also known as SPS1a~\cite{Allanach:2002nj}.

\begin{figure*}[t]
\centering
\includegraphics[height=1.0in]{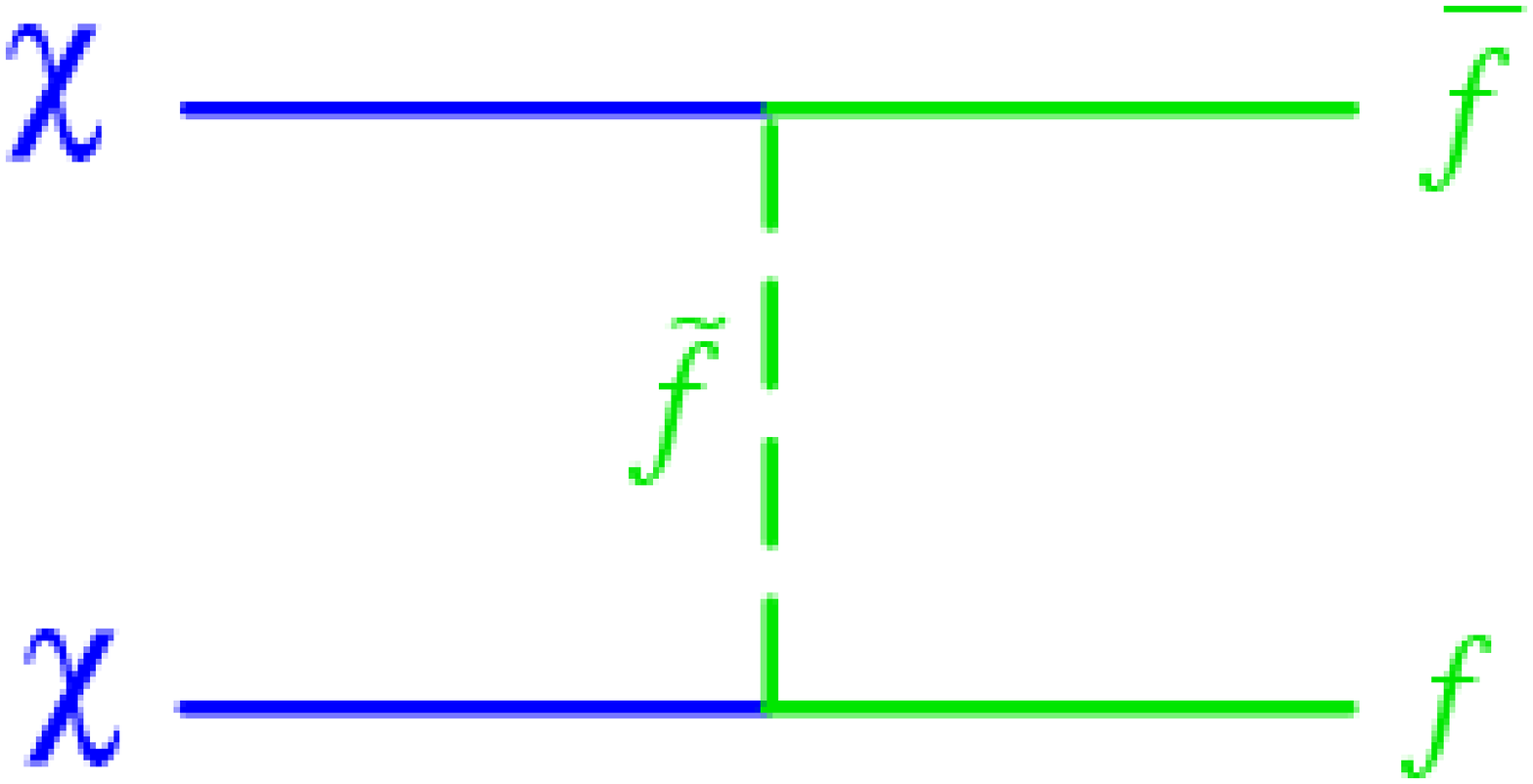}
\hspace*{0.5in}
\includegraphics[height=2.0in]{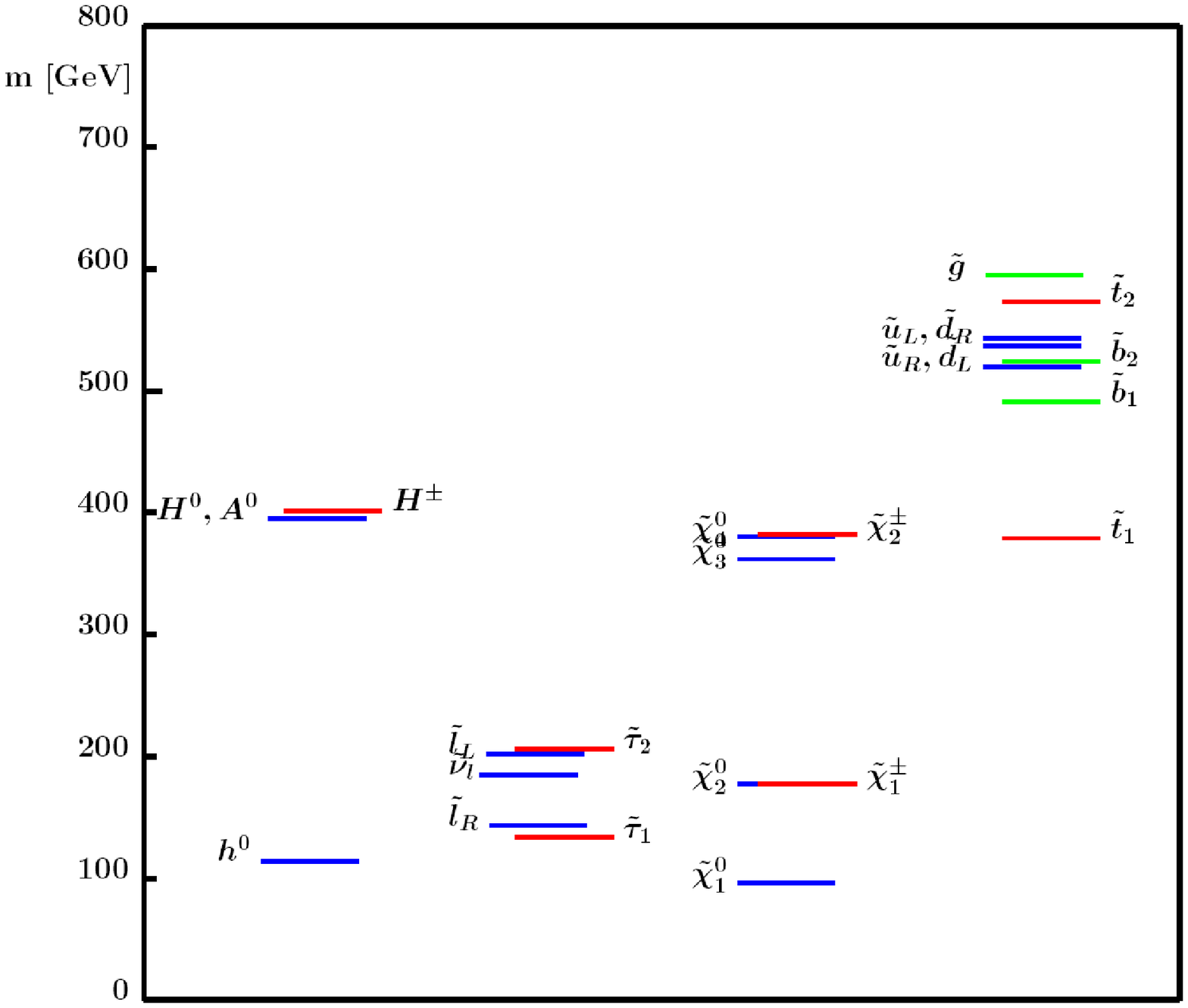}
\caption{Left: The dominant neutralino annihilation process in the
  bulk region.  Right: The superpartner spectrum at LCC1, a
  representative model in the bulk region~\cite{Allanach:2002nj}.}
\label{fig:bulk_feyn}
\end{figure*}

In the bulk region, neutralinos annihilate dominantly through $\chi
\chi \to f \bar{f}$ through a $t$-channel scalar $\tilde{f}$, as shown
in \figref{bulk_feyn}.  To achieve the correct relic density, this
process must be efficient, requiring light sfermions and, since the
neutralino must be the lightest supersymmetric particle (LSP), light
neutralinos.  These characteristics are exhibited in the superpartner
spectrum of LCC1, shown in \figref{bulk_feyn}.  It is noteworthy that
in this case, cosmology provides a strong motivation for light
superpartners within the reach of a 500 GeV ILC.

To determine the relic density at LCC1, all of the supersymmetry
parameters entering annihilation processes, including those shown in
\figref{bulk_feyn} and others, must be determined to high accuracy.
The LCC1 superpartner spectrum makes possible many high precision
measurements at the LHC.  LCC1 (SPS1a) is in significant respects a
``best case scenario'' for the LHC.  The implications of these
measurements for cosmology will be summarized below.

The LHC results may be improved at the ILC.  For example, superpartner
masses may be determined with extraordinary precision through
kinematic endpoints and threshold scans, as shown in
\figref{bulk_threshold}.  The kinematic endpoints of final state
leptons in the process $e^+ e^- \to \tilde{l}^+ \tilde{l}^- \to l^+
l^- \chi \chi$ determine both $\tilde{l}$ and $\chi$ masses.  Slepton
masses may also be determined through threshold scans.  Threshold
scans provide even higher precision, and may actually {\em save}
luminosity.  This is the case, for example, for selectron mass
determinations through $e^-e^-$ threshold scans, where precisions of
tens of MeV may be obtained with 1 to $10~\ifb$ of integrated
luminosity~\cite{Feng:1998ud,Feng:2001ce,Freitas:2003yp}.  More
generally, the required measurements exploit the full arsenal of the
ILC, from its variable beam energy, to its polarized beams, to the
$e^-e^-$ option.  The results of one study are summarized in
\figref{bulk_table}.

\begin{figure*}[t]
\centering
\includegraphics[height=2.0in]{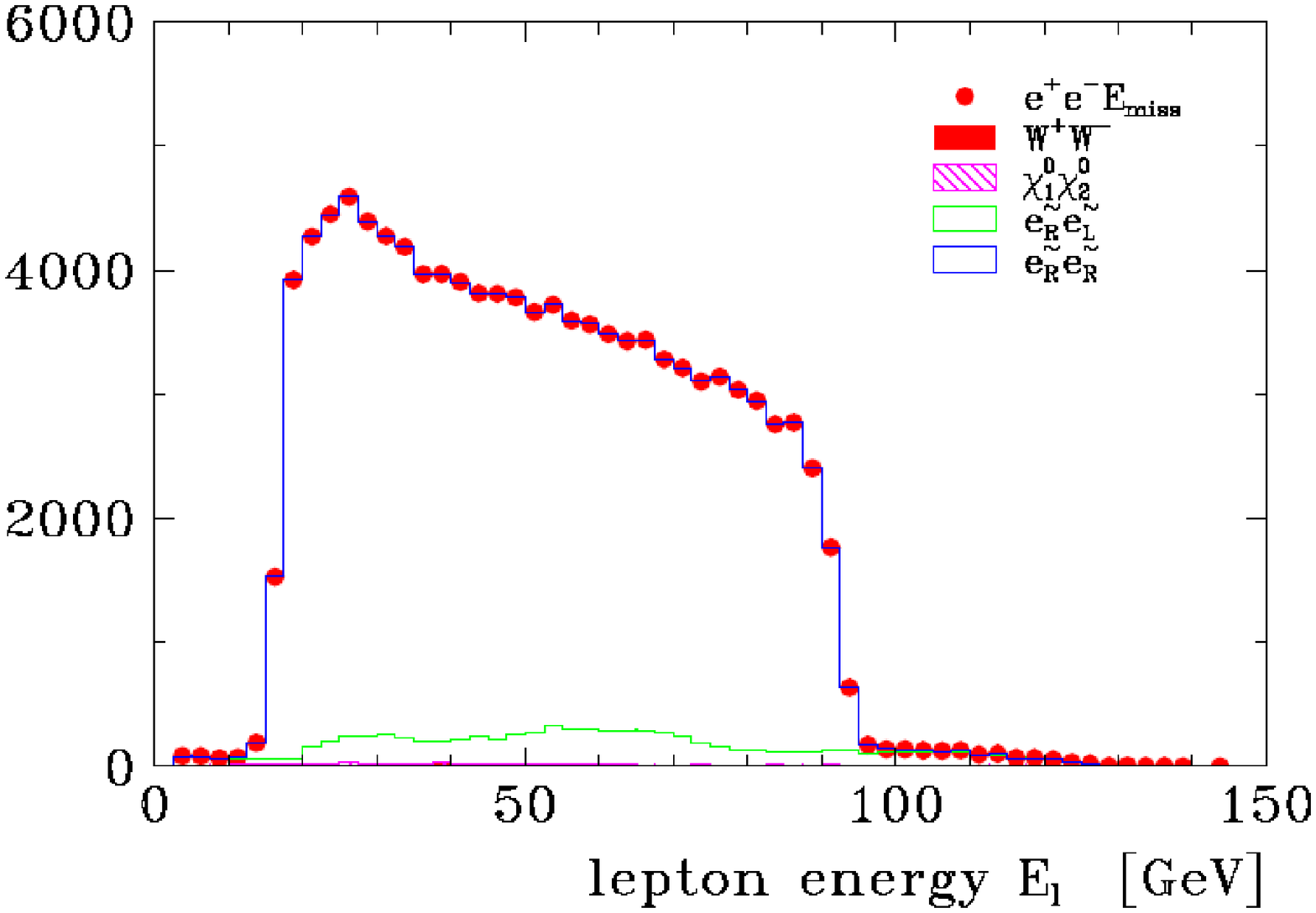}
\hspace*{0.5in}
\includegraphics[height=2.0in]{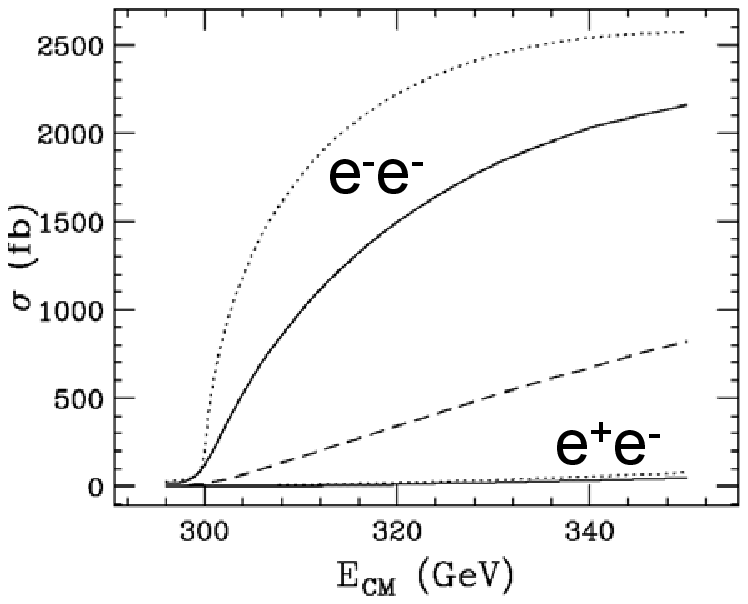}
\caption{Left: The kinematic endpoints of lepton energies from $e^+
e^- \to \tilde{l}^+ \tilde{l}^- \to l^+ l^- \chi \chi$ provide precise
determinations of slepton and neutralino
masses~\cite{Weiglein:2004hn}.  Right: Threshold scans may also be
used to determine slepton masses.  In the case of selectron masses,
$e^-e^-$ threshold scans provide higher precision and simultaneously
{\em save} luminosity~\cite{Feng:2001ce}.}
\label{fig:bulk_threshold}
\end{figure*}

\begin{figure*}[t]
\centering
\includegraphics[height=2.0in]{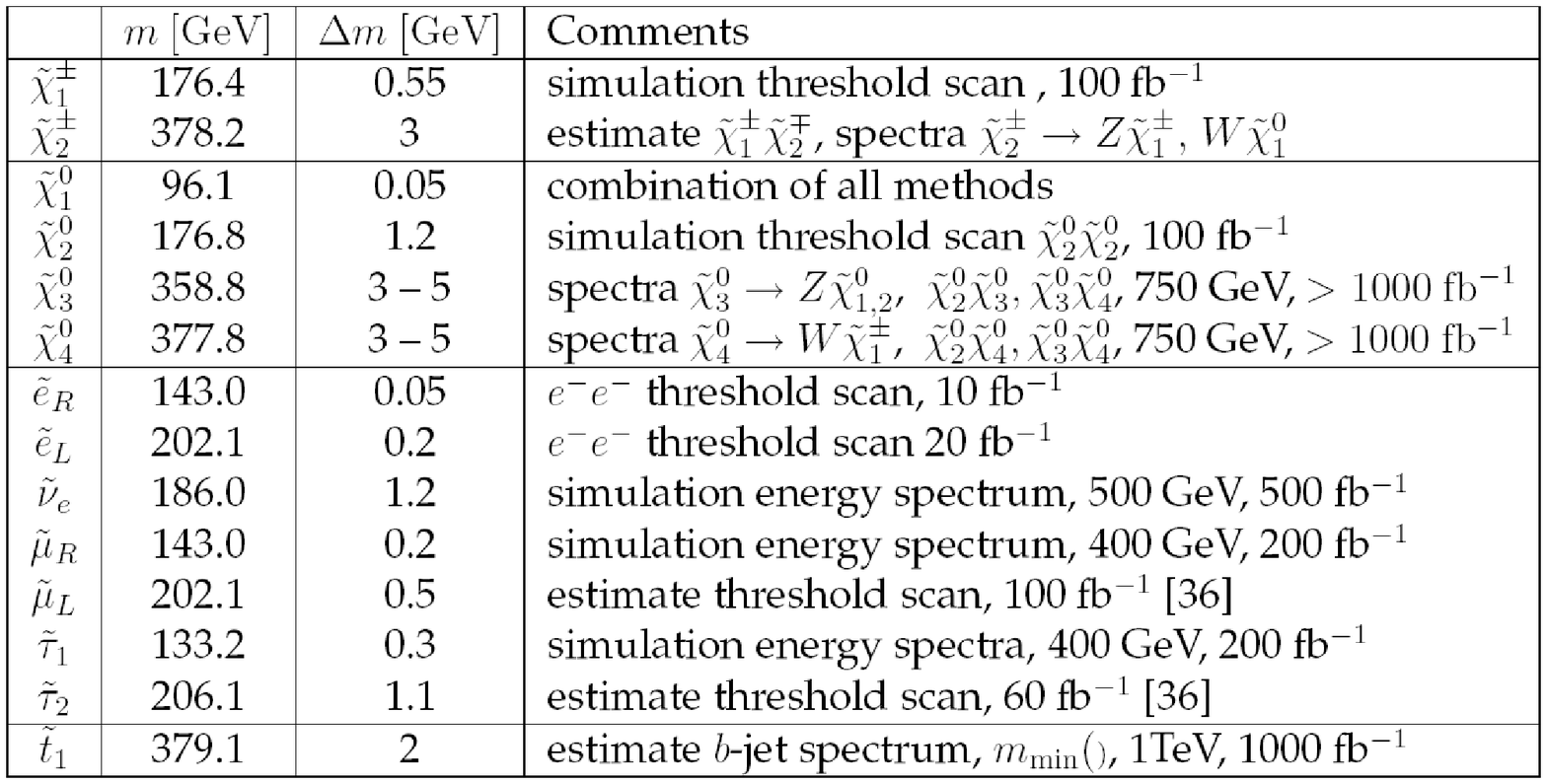}
\caption{The precision with which supersymmetry parameters may be
  determined at the ILC~\cite{Weiglein:2004hn}.}
\label{fig:bulk_table}
\end{figure*}

The neutralino thermal relic density may be determined by combining
the precise determination of all relevant supersymmetry parameters and
also verifying the insensitivity of the relic density to all other
parameters.  The results depend somewhat on the prescription one uses
to combine these data.  One approach is to choose points in parameter
space at random, weighting each with a Gaussian distribution for each
observable.  The relic density allowed region is then identified as
the symmetric interval around the central value that contains 68\% of
the weighted probability.

The result of applying this method with 50,000 scan points around LCC1
is shown in \figref{bulk_omega}.  The result is that the ILC may
determine the thermal relic density to a fractional uncertainty of
\begin{equation}
\text{LCC1 (preliminary):} \ 
\frac{\Delta (\Omegachi h^2)}{\Omegachi h^2} = 2.2\% 
\quad [ \Delta (\Omegachi h^2) = 0.0042 ] \ .
\end{equation}
The current constraint from WMAP, as well as projected future
constraints on cosmological observations from the Planck satellite
(both scaled up to the LCC1 central value for $\Omegachi$) are also
shown.  WMAP and Planck provide no information about the mass of the
dark matter particle.

\begin{figure*}[t]
\centering
\includegraphics[height=3.0in]{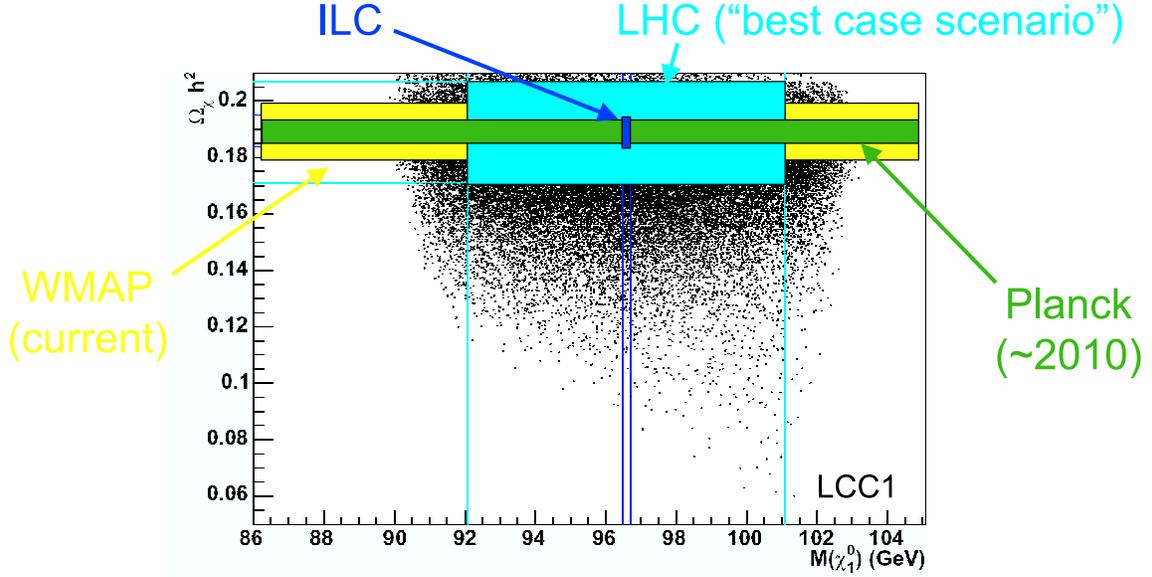}
\caption{Constraints in the $(\mchi, \Omegachi)$ plane from the ILC
  and LHC, with constraints on $\Omegachi$ from the WMAP and Planck
  satellite experiments.  The 50,000 scan points used to determine the
  ILC constraint are also shown (see text)~\cite{whitepaper}. Note
  that the distribution of scan points is much broader than the final
  ILC constrained region; out-lying points have very little
  probability weight.}
\label{fig:bulk_omega}
\end{figure*}

\paragraph{Focus Point Region}

In the focus point region, one may choose the representative model
\begin{equation}
\text{LCC2:} \ (m_0, M_{1/2}, A_0, \tan\beta) = 
(3280~\gev, 300~\gev, 0, 10) \quad
  [\mu > 0, m_{3/2} > m_{\text{LSP}}, m_t = 175~\gev] \ . 
\end{equation}
In focus point supersymmetry~\cite{Feng:1999hg,Feng:1999mn}, squarks
and sleptons are very heavy, and so the diagrams that are dominant in
the bulk region are suppressed.\footnote{As a result of this property,
models like focus point supersymmetry may be challenging for
supersymmetry discovery and study at the LHC~\cite{Baer:2005ky}.}
Nevertheless, the desired relic density may be
achieved~\cite{Feng:2000gh}, because in the focus point region, the
neutralino is not a pure Bino, but contains a significant Higgsino
component. The processes $\chi \chi \to W^+ W^-$, shown in
\figref{fp_feyn}, and $\chi \chi \to ZZ$, which are negligible in the
bulk region, therefore become efficient.  Neutralino mixing is
typically achieved when neutralinos and charginos are fairly light and
not too split in mass, and so the demands of neutralino dark matter
motivate supersymmetry with light neutralinos and charginos.

\begin{figure*}[t]
\centering \includegraphics[height=1.0in]{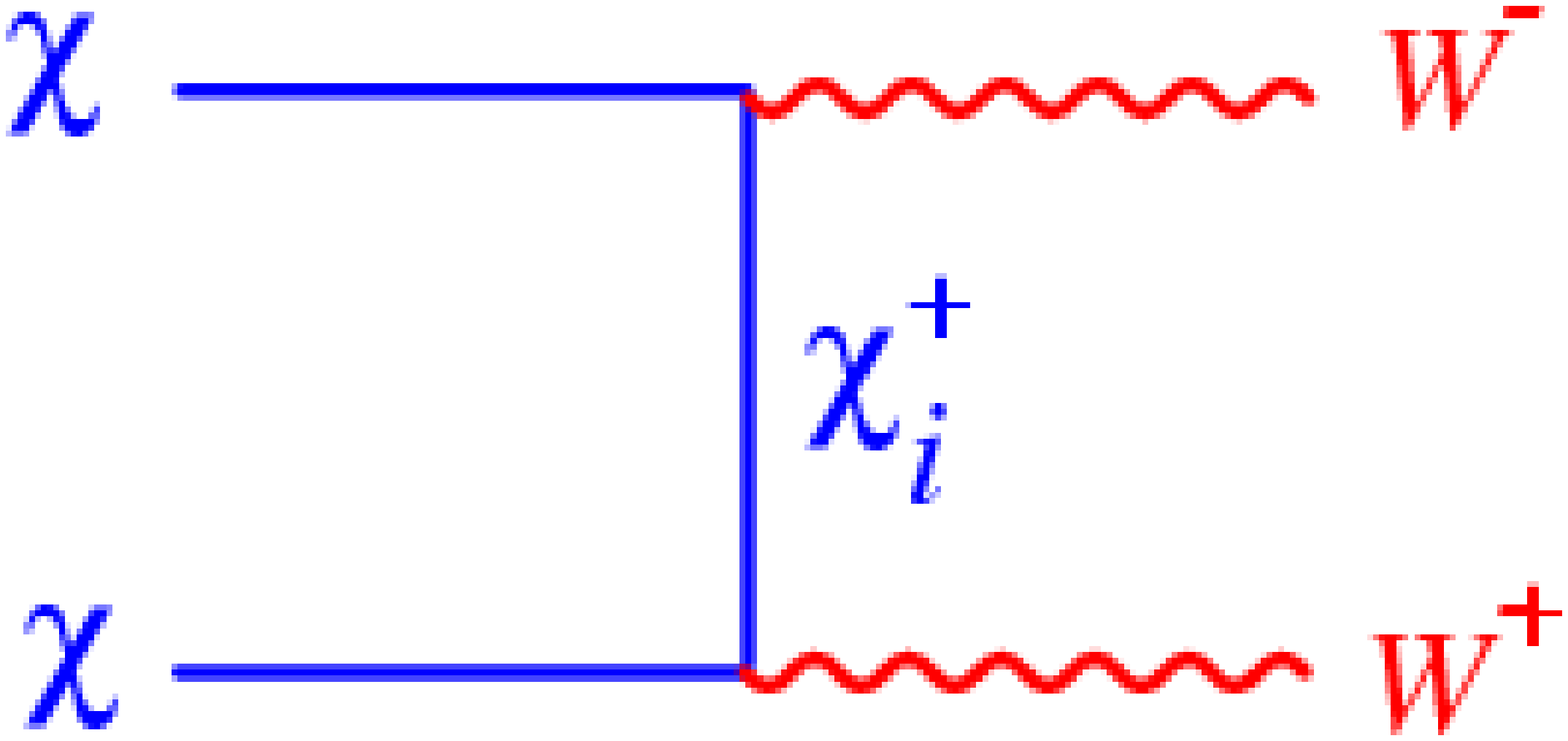}
\hspace*{0.5in}
\includegraphics[height=1.8in]{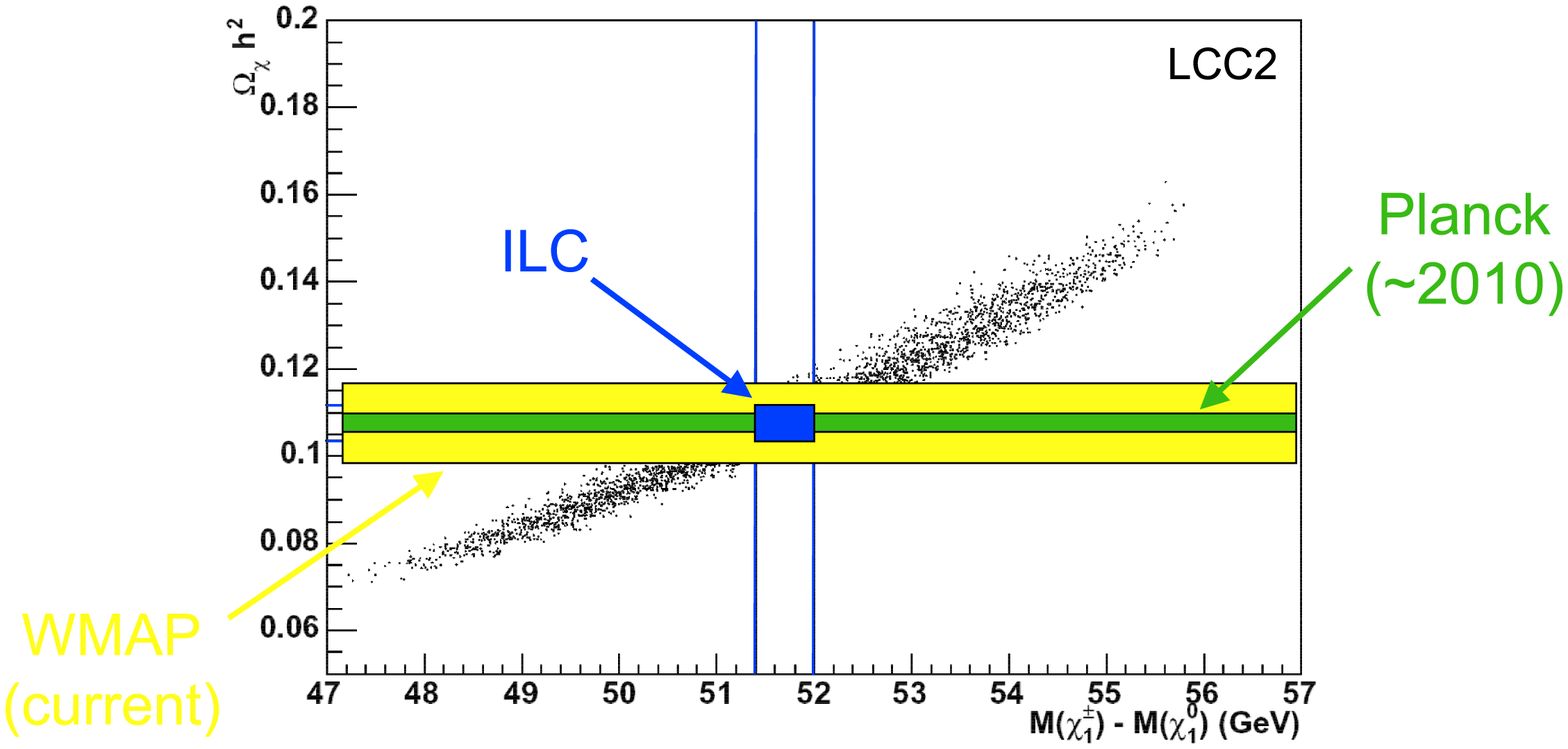}
\caption{Left: The dominant neutralino annihilation process in the
  focus point region.  Right: Constraints on the relic density from
  the ILC, WMAP, and Planck, as in
  \figref{bulk_omega}~\cite{whitepaper}.}
\label{fig:fp_feyn}
\end{figure*}

Determination of the thermal relic density in the focus point region
requires precise measurements of neutralino and chargino masses and
their mixings.  Applying the method described above for converting
collider constraints to a constraint on the thermal relic density, the
thermal relic density may be determined with fractional uncertainty
\begin{equation}
\text{LCC2 (preliminary):} \ 
\frac{\Delta (\Omegachi h^2)}{\Omegachi h^2} = 2.4\% 
\quad [ \Delta (\Omegachi h^2) = 0.0026 ] \ .
\end{equation}

\paragraph{What We Learn}

The results of \figsref{bulk_omega}{fp_feyn} imply that the ILC will
provide a part per mille determination of $\Omegachi h^2$ in these
cases, matching WMAP and even the extraordinary precision expected
from Planck.  The many possible implications of such measurements are
outlined in the flowchart of \figref{flowchart}.

\begin{figure*}[t]
\centering
\includegraphics[height=3.0in]{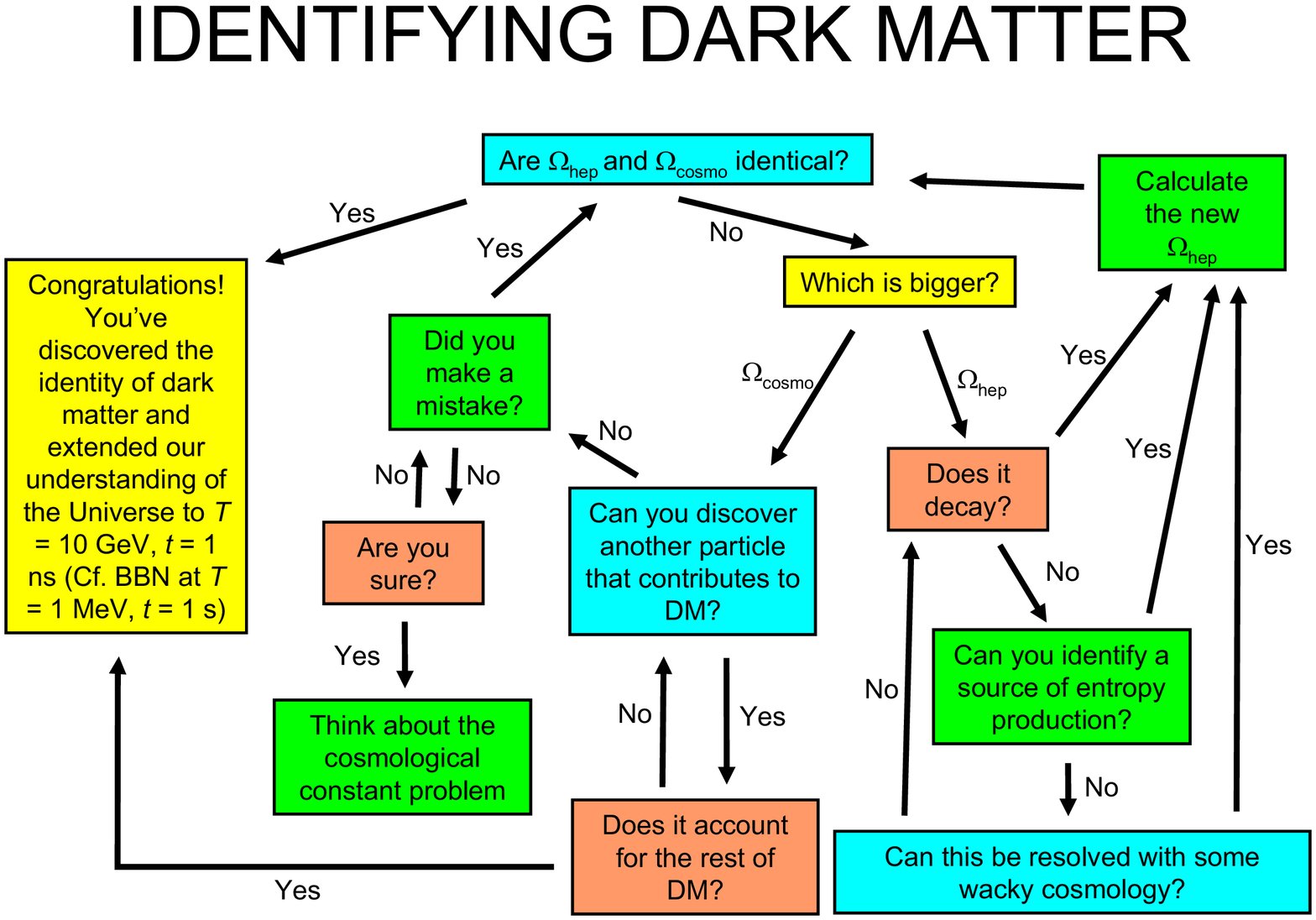}
\caption{Flowchart illustrating the possible implications of comparing
  $\Omega_{\text{hep}}$, the predicted dark matter thermal relic
  density determined from high energy physics, and
  $\Omega_{\text{cosmo}}$, the actual dark matter relic density
  determined by cosmological observations.}
\label{fig:flowchart}
\end{figure*}

Consistency of the ILC and WMAP/Planck measurements at the part per
mille level would provide strong evidence that neutralinos are
absolutely stable and form all of the non-baryonic dark matter.  Such
a result would at last provide convincing evidence that we have
produced dark matter at colliders and that we have identified its
microphysical properties.  It would be a landmark success of the
particle physics/cosmology connection, and would give us confidence in
our understanding of the Universe back to neutralino freeze out at
$t\sim 10^{-8}~\s$, eight orders of magnitude earlier than can
currently be claimed.

On the other hand, inconsistency would lead to a Pandora's box of
possibilities, all with important implications.  If the thermal relic
density determined from high energy physics is smaller than what is
required cosmologically, these high precision measurements imply that
neutralinos are at most only one component of cold, non-baryonic dark
matter.  On the other hand, if the thermal relic density determined at
colliders is too large, these measurements imply that neutralinos must
decay (perhaps to superWIMPs --- see below), or that the neutralino
thermal relic density is diluted by entropy production or some other
effect after freeze out.

The implications of LHC precision measurements for the relic density,
determined in the way discussed above, are also shown in
\figref{bulk_omega}.  The LHC precision in the LCC1 scenario is
extraordinary and unusual; for other scenarios, the LHC is unlikely to
determine $\Omegachi$ to better than one or more orders of magnitude.
At the same time, even in this ``best case scenario,'' the LHC
determination of $\Omegachi$ leaves open many possibilities.  For
example, comparison of the LHC result with WMAP/Planck cannot
differentiate between a Universe with only neutralino dark matter and
a Universe in which dark matter has two components, with neutralinos
making up only 80\%.  Such scenarios are qualitatively distinct, in
the sense that the possibility of another component with such
significant energy density can lead to highly varying conclusions
about the contents of the Universe and the evolution of structure that
formed the galaxies we see today.

\subsubsection{Mapping the WIMP Universe}

WIMPs may appear not only at colliders, but also in dark matter
searches.  Direct dark matter search experiments look for the recoil
of WIMPs scattering off highly shielded detectors.  Indirect dark
matter searches look for the products, such as positrions, gamma rays
or neutrinos, of WIMPs annihilating nearby, such as in the halo, the
galactic center, or the core of the Sun.

If WIMPs are discovered at colliders and their thermal relic densities
are determined to be cosmologically significant, it is quite likely
that they will also be discovered through direct and indirect dark
matter search experiments.  The requirement of the correct relic
density implies that WIMP annihilation was efficient in the early
Universe.  This suggests efficient annihilation now, corresponding to
significant indirect detection rates, and efficient scattering now,
corresponding to significant direct detection rates.  This rough
correspondence is illustrated in \figref{crossing}.

\begin{figure*}[t]
\centering
\includegraphics[height=2.0in]{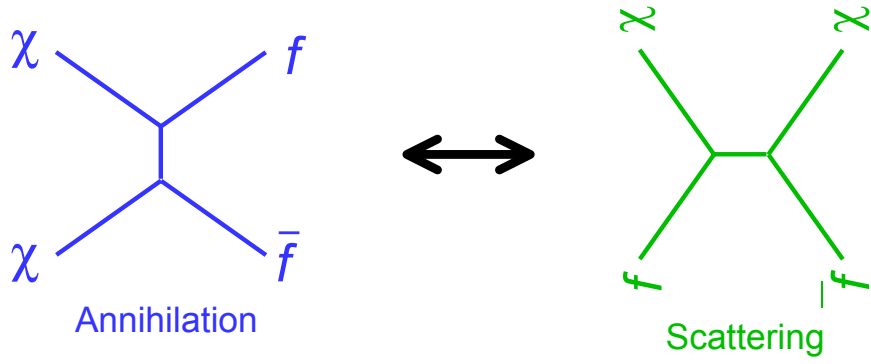}
\caption{Efficient annihilation, corresponding to large indirect
  detection rates, is related to efficient scattering, corresponding
  to large direct detection rates.}
\label{fig:crossing}
\end{figure*}

Direct and indirect dark matter detection rates are subject to
uncertainties from both particle physics, through the microphysical
properties of dark matter, and astrophysics, through the spatial and
velocity distributions of dark matter.  If completed, the research
program described in \secref{relic} to pin down the properties of
WIMPs will effectively remove particle physics uncertainties.  Dark
matter search experiments then become probes of dark matter
distributions.

As an example, consider direct detection.  Theoretical predictions of
direct detection rates are given in \figref{direct_detection}.  As is
typically done in particle physics studies, a simple dark matter halo
profile is assumed throughout this figure.  The enormous variation in
rates results from particle physics uncertainties alone.  LHC and ILC
studies will reduce this uncertainty drastically.  For example, for
LCC2, the dark matter mass will be determined to a GeV at the ILC, and
the cross section for neutralino-proton scattering will be determined
to $\Delta \sigma / \sigma \alt 10\%$~\cite{whitepaper}.  This
constraint is shown in \figref{direct_detection}, where the
uncertainties are smaller than the extent of the $\star$ plotting
symbol.

\begin{figure*}[t]
\centering
\includegraphics[height=3.0in]{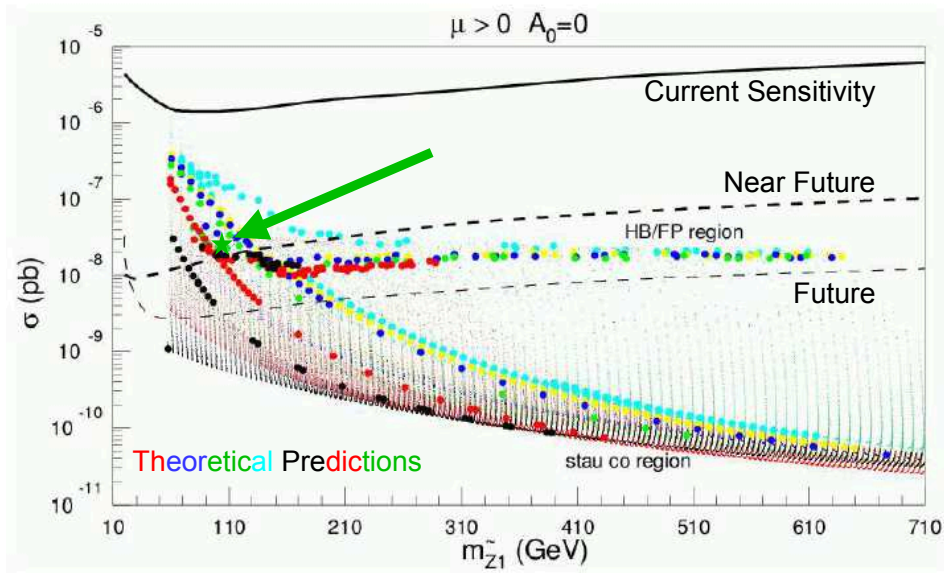}
\caption{Theoretical predictions for the direct detection
neutralino-proton scattering cross section $\sigma$, as a function of
neutralino mass $\mchi$, for various mSUGRA models
(dots)~\cite{Baer:2003jb}, and the prediction of LCC2 ($\star$).  ILC
studies will constrain the values of $\sigma$ and $\mchi$ to be
smaller than the extent of the $\star$ plotting
symbol~\cite{whitepaper}.}
\label{fig:direct_detection}
\end{figure*}

Once collider constraints effectively remove microphysical
uncertainties, the direct detection rates give us information about
the local dark matter density and velocity profile.  In a similar way,
indirect detection rates will provide additional complementary
information.  For example, experiments such as HESS and GLAST may
detect photons dark matter annihilation in the galactic center.  Such
rates are sensitive to the halo profile at the galactic center, a
quantity of great interest at present.  The synergy between collider
experiments and these dark matter experiments will constrain the phase
space distribution of WIMP dark matter in the Universe, with important
implications for the formation and evolution of structure.

\subsection{SuperWIMPs \label{sec:swimp}}

In superWIMP scenarios~\cite{Feng:2003xh}, a WIMP freezes out as
usual, but then decays to a stable dark matter particle that interacts
{\em superweakly}, as shown in \figref{freezeout_swimp}. The
prototypical example of a superWIMP is a weak-scale gravitino produced
non-thermally in the late decays of a supersymmetric WIMP, such as a
neutralino, charged slepton, or sneutrino~\cite{Feng:2003xh,%
Ellis:2003dn,Feng:2004zu,Roszkowski:2004jd}.  Additional examples
include axinos~\cite{axinos} and quintessinos~\cite{Bi:2003qa} in
supersymmetry, Kaluza-Klein graviton and axion states in models with
universal extra dimensions~\cite{Feng:2003nr}, and stable particles in
models that simultaneously address the problem of baryon
asymmetry~\cite{Kitano:2005ge}.  SuperWIMPs have all of the virtues of
WIMPs.  They exist in the same well-motivated frameworks and are
stable for the same reasons.  In addition, in the natural case that
the decaying WIMP and superWIMP have comparable masses, superWIMPs
also are naturally produced with relic densities of the desired order
of magnitude.

\begin{figure*}[t]
\centering \includegraphics[height=2.5in]{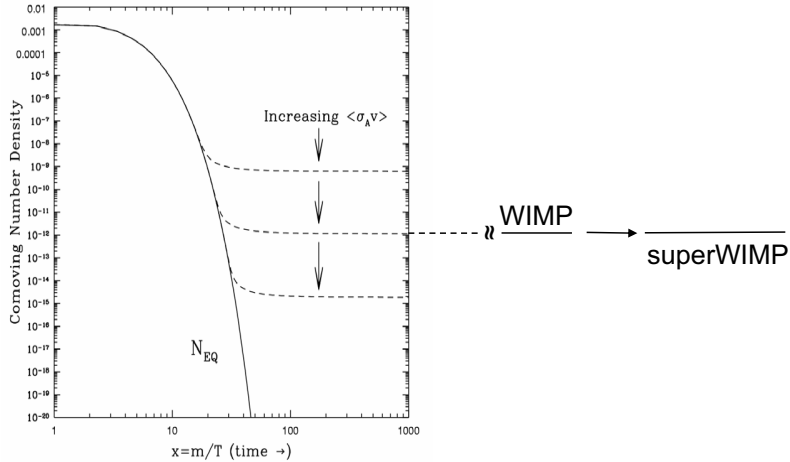}
\caption{In superWIMP scenarios, a WIMP freezes out as usual, but then
decays to a superWIMP, a superweakly-interacting particle that forms
dark matter.}
\label{fig:freezeout_swimp}
\end{figure*}

The study of superWIMP dark matter at colliders has common elements
with the study of WIMPs, but with key differences.  It may also be
divided into three (overlapping) stages:
\begin{enumerate}
\item SuperWIMP Candidate Identification.  Is there evidence for late
  decays to superWIMPs from collider studies?
\item SuperWIMP Relic Density Determination. What are the superWIMP
  candidates' predicted relic densities?  Can they be significant
  components or all of dark matter?  What are their masses, spins, and
  other quantum numbers?
\item Mapping the SuperWIMP Universe. Combined with other
  astrophysical and cosmological results, what can collider studies
  tell us about astrophysical questions, such as the distribution of
  dark matter in the Universe?
\end{enumerate}

For Stage 1, collider evidence for superWIMPs may come in one of two
forms.  Collider experiments may find evidence for charged, long-lived
particles.  Given the stringent bounds on charged dark matter, such
particles presumably decay, and their decay products may be
superWIMPs.  Alternatively, colliders may find seemingly stable WIMPs,
but the WIMP relic density studies described in \secref{relic} may
favor a relic density that is too large, providing evidence that WIMPs
decay.  These two possibilities are not mutually exclusive.  In fact,
the discovery of charged long-lived particles with too-large predicted
relic density is a distinct possibility and would provide strong
motivation for superWIMP dark matter.

In the following subsections, we will explore how well the LHC and ILC
may advance Stages 2 and 3.

\subsubsection{Relic Density Determination}

SuperWIMPs are produced in the late decays of WIMPs.  Their number
density is therefore identical to the WIMP number density at freeze
out, and the superWIMP relic density is
\begin{equation}
\Omega_{\text{sWIMP}} = \frac{m_{\text{sWIMP}}}{m_{\text{WIMP}}}
\Omega_{\text{WIMP}} \ .
\label{swimp_omega}
\end{equation}
To determine the superWIMP relic density, we must therefore determine
the superWIMP's mass.  This is not easy, since the WIMP lifetime may
be very large, implying that superWIMPs are typically produced long
after the WIMPs have escaped collider detectors.

For concreteness, consider the case of supersymmetry with a stau
next-to-lightest supersymmetric particle (NLSP) decaying to a
gravitino superWIMP.\footnote{If superWIMPs are produced in sufficient
numbers to be much of the dark matter, neutralino NLSPs are heavily
disfavored, as their late decays invariably violate constraints from
BBN and the CMB~\cite{Feng:2003xh,Feng:2004zu,Roszkowski:2004jd}.}
The stau is a particle with mass near $\mweak \sim 100~\gev$ decaying
through gravitational interactions suppressed by the (reduced) Planck
mass $\mstar \simeq 2.4 \times 10^{18}~\gev$.  On dimensional grounds,
we therefore expect its lifetime to be $\mstar^2/\mweak^3$.  More
precisely, we find
\begin{equation}
 \tau(\tilde{\tau} \to \tau \tilde{G}) = 48\pi \mstar^2
 \frac{m_{\tilde{G}}^2}{m_{\tilde{\tau}}^5}
 \left[1 -\frac{m_{\tilde{G}}^2}{m_{\tilde{\tau}}^2} \right]^{-4} 
 \sim 10^4 - 10^8~\s \ .
\label{sfermionwidth}
\end{equation}
This is outlandishly long by particle physics standards.  This
gravitino superWIMP scenario therefore implies that the signal of
supersymmetry at colliders will be meta-stable sleptons with lifetimes
of days to months.  Such particles will produce highly-ionizing tracks
that should be spectacularly obvious at the
LHC~\cite{Drees:1990yw,Goity:1993ih,Nisati:1997gb,Feng:1997zr}.  

At the same time, because some sleptons will be slowly moving and
highly-ionizing, they may be trapped and
studied~\cite{Feng:2004yi,Hamaguchi:2004df,DeRoeck:2005bw}.  As an
example, sleptons may be trapped in water tanks placed outside
collider detectors.  These water tanks may then be drained
periodically to underground reservoirs where slepton decays may be
observed in quiet environments.  This possibility has been studied in
Ref.~\cite{Feng:2004yi} and is illustrated in \figref{trap_ILC}.

\begin{figure*}[t]
\centering
\includegraphics[height=2.0in]{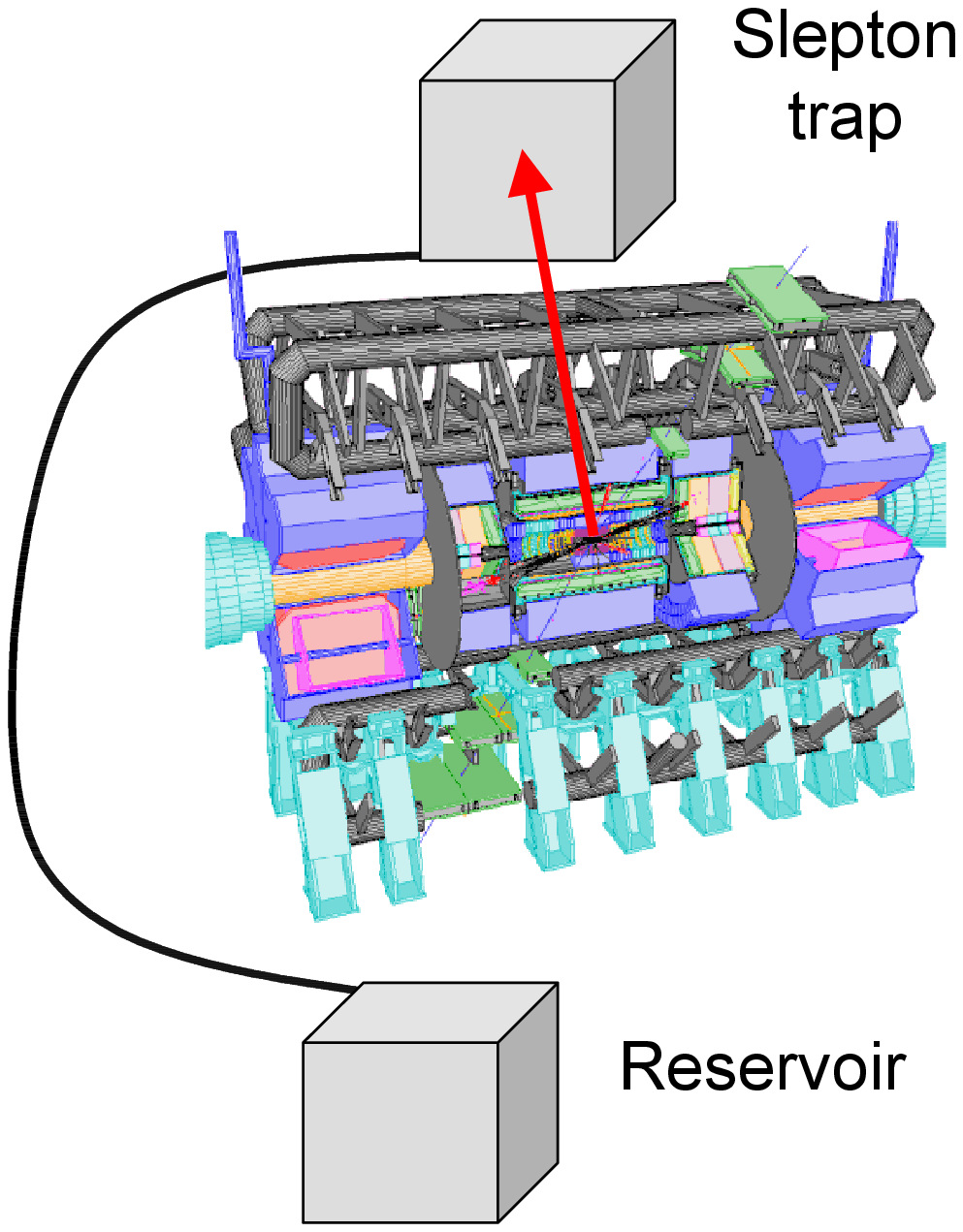}
\hspace*{1.0in}
\includegraphics[height=2.0in]{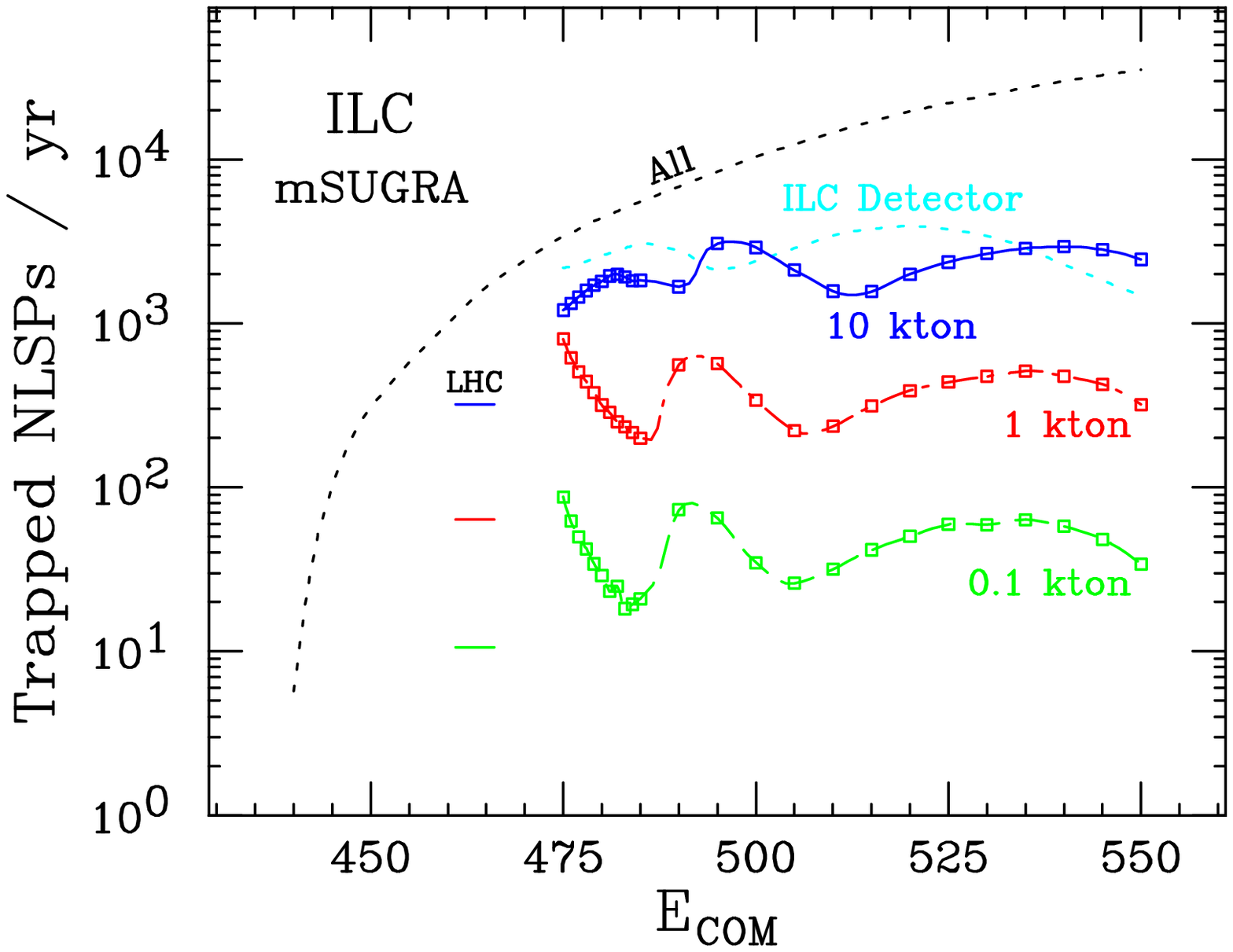}
\caption{Left: Configuration for slepton trapping in gravitino
superWIMP scenarios.  Right: The number of sleptons trapped per year
at the ILC in 10 kton (solid), 1 kton (dot-dashed), and 0.1 kton
(dashed) water traps.  The total number of sleptons produced is also
shown (upper dotted) along with the number of sleptons trapped in the
ILC detector (lower dotted).  The trap shape and placement have been
optimized and a luminosity of $300~\ifb/\yr$ is assumed.  The
underlying model is minimal supergravity with $\mgaugino = 600~\gev$,
$m_0 = 0$, $A_0 = 0$, $\tan\beta = 10$, and $\mu > 0$.  The LHC
results for this model are as indicated~\protect\cite{Feng:2004yi}.}
\label{fig:trap_ILC}
\end{figure*}

How many sleptons may be stopped in a reasonably sized trap?  The
answer is highly model-dependent.  The results for one model with 219
GeV sleptons is shown in \figref{trap_ILC}.  At the LHC, hundreds of
sleptons may be caught each year in a 10 kton trap, assuming a
luminosity of $100~\ifb/\yr$.  These LHC results may be improved
significantly if long-lived NLSP sleptons are kinematically accessible
at the ILC.  For the identical case with 219 GeV sleptons, ${\cal
O}(1000)$ sleptons may be trapped each year in a 10 kton trap at the
ILC, assuming $300~\ifb/\yr$.  By considering the slightly more
general possibility of placing lead or other dense material between
the ILC detector and the slepton trap, a further enhancement of an
order of magnitude may be possible, allowing up to ${\cal O}(10^4)$
sleptons to be trapped per ILC year.  These ILC results are made
possible by the ability to tune the beam energy to produce slow NLSPs.
The ability to prepare initial states with well-known energies and the
flexibility to tune this energy are well-known advantages of the ILC.
Here, these features are exploited in a qualitatively new way to
produce slow sleptons that are easily captured.

If thousands of sleptons are trapped, the slepton lifetime may be
determined to the few percent level simply by counting the number of
slepton decays as a function of time.  The slepton mass will be
constrained by analysis of the collider event kinematics. A per cent
level measurement of the slepton lifetime given in
\eqref{sfermionwidth} therefore implies a high precision measurement
of the gravitino mass, and therefore a determination of the gravitino
relic density through \eqref{swimp_omega}.  As with the case of WIMPs,
consistency at the percent level with the observed dark matter relic
density will provide strong evidence that dark matter is indeed
composed of gravitino superWIMPs.

SuperWIMP quantum numbers and couplings may also be determined through
collider studies~\cite{Buchmuller:2004rq,Feng:2004gn}, although, as
indicated above, these will typically be determined after or at the
same time as the relic density determination, in contrast to the case
of WIMPs.  For example, an alternative method to determine the
gravitino mass is to measure the energy of slepton decay products.
This provides a consistency check of the mass determination described
above.  Alternatively, these two methods, when combined, determine not
only $m_{\gravitino}$, but also the Planck mass $\mstar$.  Given
enough events, the gravitino spin may also be constrained to be
3/2~\cite{Buchmuller:2004rq}.  The spin and couplings of the gravitino
may therefore be determined, showing that the superWIMP is in fact the
superpartner of the graviton and that nature is locally
supersymmetric.

\subsubsection{Mapping the SuperWIMP Universe}

Collider studies of superWIMPs will have significant implications for
the phase space distribution of dark matter.  In fact, the discovery
of superWIMPs may resolve current discrepancies and shed light on
important and controversial issues in structure formation.

In the standard cosmology, dark matter is assumed to be cold, as is
the case with WIMP dark matter.  Cold dark matter is remarkably
successful in explaining the observed large scale structure down to
length scales of $\sim 1~\Mpc$.  Despite its considerable virtues,
however, cold dark matter appears to face difficulty in explaining the
observed structure on length scales $\alt 1~\Mpc$. Numerical
simulations assuming cold dark matter predict, for example, overdense
cores in galactic halos~\cite{Moore:1994yx} and too many dwarf
galaxies in the Local Group~\cite{Klypin:1999uc}.

These problems may be resolved by superWIMP dark
matter~\cite{Sigurdson:2003vy,Profumo:2004qt,Kaplinghat:2005sy,%
Cembranos:2005us,Jedamzik:2005sx}.  SuperWIMPs are produced with
relativistic velocities at late times, as we have seen.  They
therefore exhibit properties typically associated with warm dark
matter, suppressing power on small scales and potentially resolving
the problems of cold dark matter mentioned above.  The discovery of
superWIMP dark matter and the determination of NLSP and superWIMP
masses and other relevant parameters at colliders would therefore
change fundamentally our understanding of how galaxies were formed and
provide a new framework for understanding halo profiles and the
distribution of dark matter.

Decays that produce superWIMPs also typically release electromagnetic
and hadronic energy.  This energy may modify the light element
abundances predicted by standard BBN~\cite{Feng:2003xh,Cyburt:2002uv,%
Jedamzik:2004er,Kawasaki:2004yh,Ellis:2005ii} or distort the black
body spectrum of the
CMB~\cite{Feng:2003xh,Hu:1993gc,Fixsen:1996nj,Lamon:2005jc}.  Collider
studies will be able to determine how much energy is released and at
what time, providing still more information with important
consequences astrophysics.

\section{BARYOGENESIS}

In addition to sharpening the problems of dark matter, recent
cosmological observations have bolstered constraints from BBN to
provide tight bounds on the amount of baryons in the Universe.  Such
progress highlights our ignorance of the origin of the asymmetry
between matter and anti-matter and the mechanism of baryogenesis.

The generation of a net baryon number requires $B$ violation, $CP$
violation, and a period of departure from thermal equilibrium.  There
are many scenarios for realizing these conditions, with some of the
most attractive proposing baryogenesis or leptogenesis at the GUT
scale.  In such cases, the investigation of baryogenesis and a
quantitative explanation for $\OmegaB$ will likely be beyond the reach
of proposed colliders.

At the same time, all three conditions for baryogenesis may be
realized at the weak scale~\cite{Trodden:1998ym}. For example, a
baryon excess of the desired size may be generated with weak-scale
supersymmetry~\cite{Carena:1997gx,Carena:2002ss}.  This scenario is at
present highly constrained, but it provides a concrete example of what
is required for the quantitative investigation of baryogenesis at
colliders.

In the MSSM, weak-scale baryogenesis requires a Higgs boson mass $m_h
\alt 118~\gev$ and top squark mass below the top quark mass.  The
region of $(m_h, m_{\tilde{t}})$ parameter space compatible with
baryogenesis is shown in \figref{baryogenesis}.  If such a scenario is
realized in nature, then, precision Higgs and squark studies will be
possible at the ILC.  The potential of the ILC for both Higgs and top
squark studies has been carefully studied.  For example, top squark
masses, as well as left-right mixing in the top squark sector may be
tightly constrained through polarized cross section measurements.  The
results of one study are shown in \figref{baryogenesis}.  Top squark
mass measurements at the GeV level may favor or disfavor weak-scale
baryogenesis.

\begin{figure*}[t]
\centering
\includegraphics[height=2.0in]{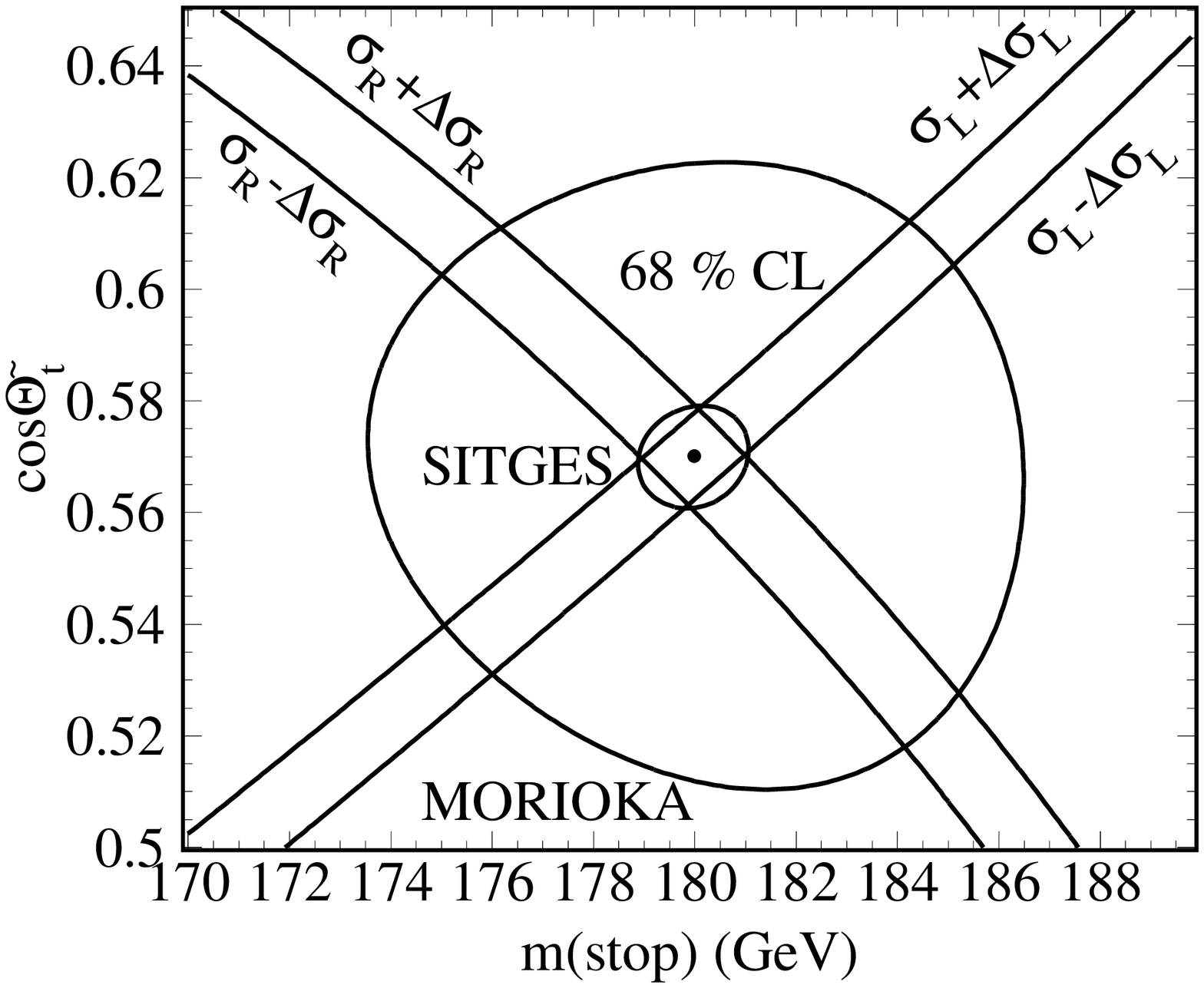}
\hspace*{0.5in}
\includegraphics[height=2.0in]{baryogenesis_bound.eps}
\caption{Left: Constraints on top squark masses and mixings from
  polarized cross section measurements at the
  ILC~\cite{Berggren:1999ss}.  Right: The region of $(m_h,
  m_{\tilde{t}})$ parameter space (unshaded) in which weak-scale
  baryogenesis is possible~\cite{Quiros:2000wk}.}
\label{fig:baryogenesis}
\end{figure*}

The full investigation of baryogenesis at colliders requires also the
measurement of CP-violating phases. Such studies have been carried
out~\cite{Ibrahim:1998je,Choi:2001ww,Barger:2001nu,Heinemeyer:2005uf,%
Feng:2002tc}.  If light top squarks are produced at the LHC and ILC,
such studies may favor or exclude weak-scale baryogenesis.  They may
even provide quantitative estimates from particle physics of
$\OmegaB$, which may be compared with the cosmological observations,
leading to implications similar to those for dark matter illustrated
in \figref{flowchart}.

\section{DARK ENERGY}

Recent observations of dark energy provide profound problems for
particle physics.  In quantum mechanics, an oscillator has zero-point
energy $\frac{1}{2} \hbar \omega$.  In quantum field theory, the
vacuum energy receives contributions of this size from each mode, and
so is expected to be $\rho_{\Lambda} \sim \int^E d^3k \frac{1}{2}
\hbar \omega \sim E^4$, where $E$ is the energy scale up to which the
theory is valid.  Typical expectations for $\rho_{\Lambda}^{1/4}$ are
therefore the weak scale or higher, whereas the observed value is
$\rho_{\Lambda}^{1/4} \sim \ \text{meV}$.  This discrepancy is the
cosmological constant problem.  Its difficulty stems from the fact
that the natural energy scale for solutions is not at high energies
yet to be explored, but at low energies that are seemingly
well-understood.

Two approaches to the cosmological constant problem are illustrated in
\figref{de_symmetry}.  In the first, something, perhaps a symmetry, is
assumed to set the cosmological constant to zero.  Additional
contributions, for example, of size $m_{\nu}^4$~\cite{Fardon:2003eh}
or $(\mweak^2/\mplanck)^4$~\cite{Kim:2002tq}, bring $\rho_{\Lambda}$
up to its observed value.  In the second approach, one assumes that
there are many possible vacua with energies densely spaced around
zero, as possibly provided by string theory~\cite{Bousso:2000xa}.  One
then hopes that anthropic arguments can explain why values of
$\rho_{\Lambda}$ near the observed value and not much larger are
favored~\cite{Weinberg:1987dv}.

\begin{figure*}[t]
\centering
\includegraphics[height=1.5in]{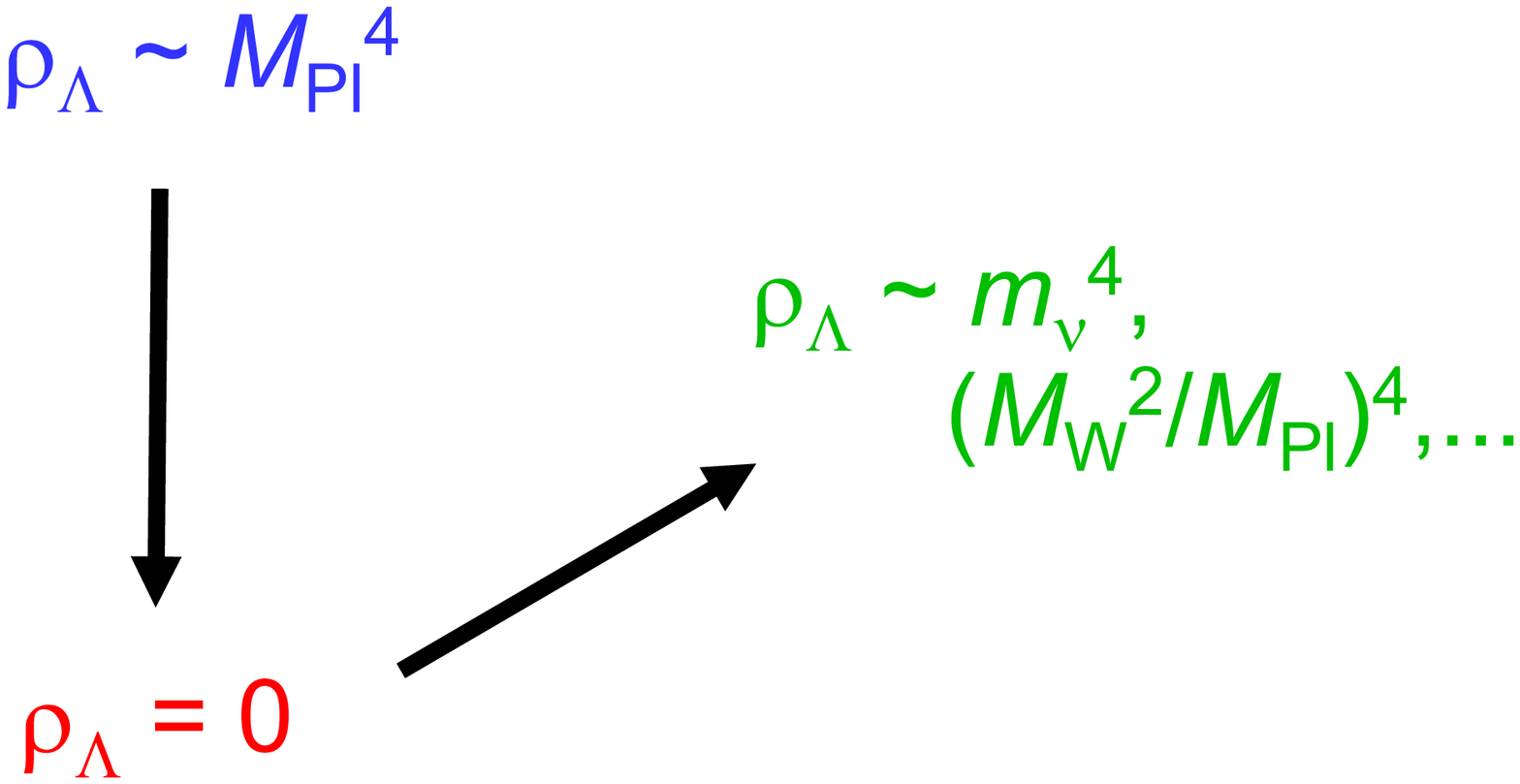}
\hspace*{0.25in}
\includegraphics[height=1.5in]{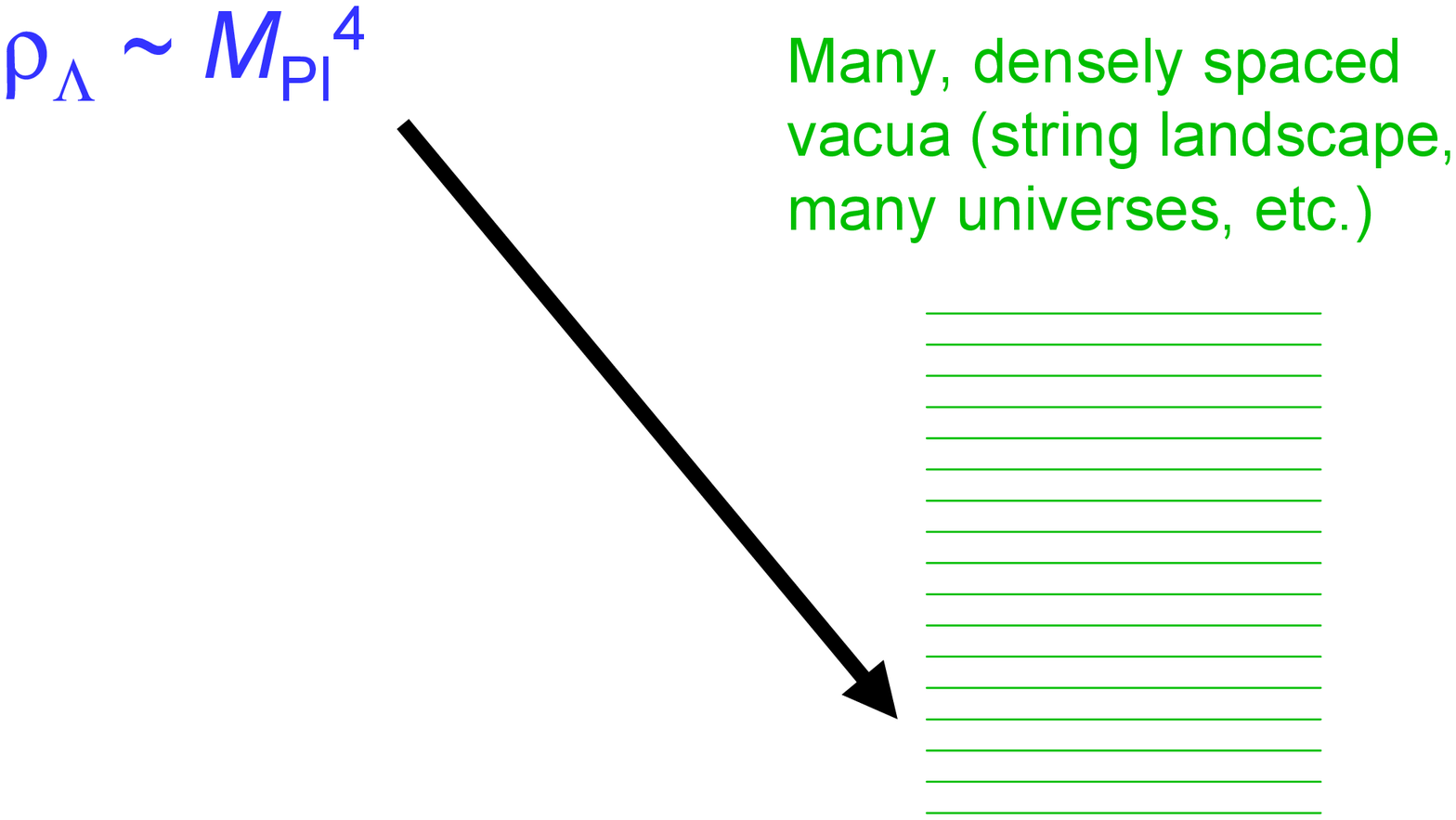}
\caption{Symmetry (left) and anthropic (right) approaches to
the cosmological constant problem. }
\label{fig:de_symmetry}
\end{figure*}

The symmetry and anthropic approaches to the cosmological constant
problem are completely different.  Their only similarity is that the
more one thinks about either of them, the more one concludes that the
other approach must be more promising.  There are other proposals,
also creative and intriguing, but so far none is particularly
compelling.  

Perhaps it is not surprising that proposed solutions seem less than
satisfactory, however.  The cosmological constant is very likely
signaling a fundamental problem with our basic ideas, much as the
black body problem signaled a problem with classical mechanics at the
previous turn-of-the-century.  The quantum revolution was decades in
the making, and we have had only a few years to puzzle over the modern
form of the cosmological constant problem.

Can upcoming colliders provide any insights? It would be pure fancy at
this stage to propose an experimental program to solve the
cosmological constant problem.  On the other hand, the LHC and ILC may
contribute in related areas, such as to our understanding of scalar
fields and vacuum energy.  For example, these colliders may
\begin{enumerate}
\item Discover a fundamental scalar particle.  Such particles are
  critical to most discussions of dark energy, from inflation to
  quintessence to the cosmological constant, but none has yet been
  observed.
\item Investigate contributions of order $\mweak^4 \sim 10^{60}
  \rho_{\Lambda}$ by mapping out the electroweak potential through
  studies of Yukawa couplings and Higgs self-couplings.
\item Investigate contributions of order $\msusy^4 \sim 10^{90}
  \rho_{\Lambda}$ by measuring the gravitino mass, perhaps as
  discussed in \secref{swimp}, and with it, the scale of supersymmetry
  breaking.
\item Investigate contributions of order $\mgut^4 \sim 10^{108}
  \rho_{\Lambda}$ by extrapolating weak scale measurements to the GUT
  scale.
\end{enumerate}
It is far from clear that these steps will lead to more compelling
explanations of dark energy.  At the same time, these steps will
constitute progress in closely related topics, and one should not
underestimate the power of experimental data to catalyze
breakthroughs. We hope that 1 will largely be accomplished when the
LHC discovers one or more Higgs bosons, whose properties will then be
precisely constrained by the LHC and ILC.  For 2, 3, and 4, to the
extent the relevant ideas are realized in nature, the ILC will likely
be an essential tool.

\section{CONCLUSIONS}

We live at a doubly exciting time in particle physics.  While
important questions related to flavor and electroweak symmetry
breaking remain, breakthroughs in cosmology have added a whole new
layer of fundamental problems requiring particle physics answers.  In
the next few years, the Tevatron and LHC will begin to explore the
weak scale. If there is new physics, experiments at these colliders
are likely to discover it, and detailed studies will likely also be
possible, especially at the LHC.

What can the ILC add?  In particle physics, extensive studies have
shown that the ILC can free experimental studies from many theoretical
assumptions and simultaneously increase the precision of measurements
of many observables.  For the cosmological questions explored here,
these same characteristics carry over.  More than that, however, in
many well-motivated examples, the increased precision at the ILC is
required to make full use of the precision now obtained in cosmology,
and the resulting synergy may lead to {\em qualitatively} new
conclusions, such as the definitive identification of dark matter.
The cosmological and astrophysics topics potentially addressed by the
ILC are wide-ranging and include dark matter and dark energy, the
mechanism of baryogenesis, Big Bang nucleosynthesis, the cosmic
microwave background, and the origin of large scale structure.  If any
of the ideas discussed here is realized in nature, the coming
revolution in particle physics will also yield profound insights about
the Universe, its contents, and its evolution.

\begin{acknowledgments}
It is a pleasure to thank the members of the ALCPG Cosmology Subgroup,
particularly my co-editors Marco Battaglia, Norman Graf, Michael
Peskin, and Mark Trodden and collaborators Jose Ruiz Cembranos,
Konstanin Matchev, Arvind Rajaraman, Bryan Smith, Shufang Su, Fumihiro
Takayama, and Frank Wilczek for many contributions to the viewpoints
and results summarized here.  The work of JLF is supported in part by
NSF CAREER grant No.~PHY--0239817, NASA Grant No.~NNG05GG44G, and the
Alfred P.~Sloan Foundation.
\end{acknowledgments}

\end{document}